\documentclass{IOS-Book-Article}

\usepackage{array}
\usepackage{multirow}
\usepackage{amsmath}
\usepackage{amsfonts}
\usepackage{amssymb}
\usepackage{mathptmx}
\usepackage{graphicx}
\usepackage{cite}
\usepackage[caption=false, subrefformat=parens, labelformat=parens]{subfig}
\usepackage[T1]{fontenc}
\usepackage[dvipsnames]{xcolor}
\usepackage[normalem]{ulem}   
\def\hb{\hbox to 10.7 cm{}}

\newcommand{\ea}{{\it et al.\/}}

\def\ket#1{\left|#1\right>}

\begin{document}

\pagestyle{headings}
\def\thepage{}

\begin{frontmatter}              

\title{From Ans\"atze to $Z$-gates: \\
a NASA View of Quantum Computing
}

\markboth{}{May 2019\hb}

\author[QuAIL]{\fnms{Eleanor G.} \snm{Rieffel}%
\thanks{Senior Researcher, QuAIL Lead. Corresponding Author: eleanor.g.rieffel@nasa.gov}},
\author[QuAIL,USRA]{\fnms{Stuart} \snm{Hadfield}},
\author[QuAIL,USRA]{\fnms{Tad} \snm{Hogg}},
\author[QuAIL,SGT]{\fnms{Salvatore} \snm{Mandr\`a}},
\author[QuAIL,USRA]{\fnms{Jeffrey} \snm{Marshall}},
\author[QuAIL,SGT]{\fnms{Gianni} \snm{Mossi}},
\author[QuAIL,UCB]{\fnms{Bryan} \snm{O'Gorman}},
\author[QuAIL,USRA]{\fnms{Eugeniu} \snm{Plamadeala}},
\author[QuAIL]{\fnms{Norm M.} \snm{Tubman}},
\author[QuAIL,USRA]{\fnms{Davide} \snm{Venturelli}},
\author[QuAIL,SGT]{\fnms{Walter} \snm{Vinci}},
\author[QuAIL,USRA]{\fnms{Zhihui} \snm{Wang}},
\author[QuAIL,SGT]{\fnms{Max} \snm{Wilson}},
\author[QuAIL,USRA]{\fnms{Filip} \snm{Wudarski}},
and
\author[QuAIL]{\fnms{Rupak} \snm{Biswas} \thanks{Director of Exploration Technology}}

\runningauthor{E. G. Rieffel et al.}
\address[QuAIL]{Quantum Artificial Intelligence Lab. (QuAIL), Exploration Technology Directorate, NASA Ames Research Center, Moffett Field, CA 94035, USA}
\address[SGT]{Stinger Ghaffarian Technologies Inc., Greenbelt, MD 20770, USA}
\address[UCB]{University of California, Berkeley, CA 94720, USA}
\address[USRA]{USRA Research Institute for Advanced Computer Science (RIACS), Mountain View, CA 94035, USA}

\begin{abstract}

For the last few years, the NASA Quantum Artificial Intelligence Laboratory (QuAIL) has been performing research to assess the potential impact of quantum computers on challenging computational problems relevant to future NASA missions. A key aspect of this research is devising methods to most effectively utilize emerging quantum computing hardware. Research questions include what experiments on early quantum hardware would give the most insight into the potential impact of quantum computing, the design of algorithms to explore on such hardware, and the development of tools to minimize the quantum resource requirements. 
We survey work relevant to these questions, with a particular emphasis on our recent work in quantum algorithms and applications, in elucidating mechanisms of quantum mechanics and their uses for quantum computational purposes, and in simulation, compilation, and physics-inspired classical algorithms. To our early application thrusts in planning and scheduling, fault diagnosis, and machine learning, we add thrusts related to robustness of communication networks and the simulation of many-body systems for material science and chemistry. We provide a brief update on quantum annealing work, but concentrate on gate-model quantum computing research advances within the last couple of years.

\end{abstract}

\end{frontmatter}
\markboth{May 2019\hb}{April 2019\hb}

\section{Introduction}

The power of quantum computation comes from encoding information in a non-classical way, enabling quantum algorithms to harness effects at the heart of quantum mechanics -- interference,  tunneling,  entanglement, measurement, many-body delocalization
-- for computational purposes. Several 
quantum algorithms are known that provably outperform the best classical algorithms \cite{RPbook,nielsen2010quantum,QAlgsZoo}. The field has matured rapidly from its birth in the early 1980s, blossoming with Shor's 1994 breakthrough discovery of polynomial-time quantum algorithms for the cryptographically important integer factoring and discrete logarithm problems \cite{shor1994algorithms}. 

While large-scale universal quantum computers are likely decades away, the next few years will see the emergence of prototype quantum processors that can run a wide variety of quantum algorithms whose simulation is beyond the reach of even the most powerful supercomputers. The advent of such hardware opens up empirical exploration of quantum algorithms far beyond what has been possible to date, including the potential to establish novel quantum heuristic algorithms. Heuristic algorithms commonly serve as the leading approaches to solving many of the world's most challenging computational problems. With the advent of more sophisticated early quantum hardware, and the empirical exploration it enables, we expect a substantial broadening of established applications of quantum computing. Target applications include optimization, machine learning, simulation of quantum chemistry and material science, and beyond. 

The earliest available quantum computing hardware were D-Wave quantum annealers, special-purpose  processors that could run one type of quantum algorithm, quantum annealing, an optimization metaheuristic.
The world is now entering the Noisy Intermediate-Scale Quantum (NISQ) \cite{preskill2018quantum} era, with prototype universal gate-model quantum processors, 
including superconducting processors at
Google \cite{boixo2018characterizing}, IBM \cite{IBM}, and Rigetti \cite{Sete16}, as well as 
more sophisticated quantum annealers.
Universal quantum processors are on the verge of ``quantum supremacy'' \cite{preskill2012quantum}, 
the ability to perform computations beyond the reach of even the largest supercomputers. However, useful quantum supremacy - the ability of a quantum processor to provide a solution to a practical problem that cannot be found by supercomputers - is still some years away. In the meantime, emerging quantum processors provide an unprecedented opportunity to explore quantum algorithms, providing the means to test quantum algorithms at sizes beyond classical simulation. 
A critical question for the NISQ era is what should be run on these early devices to give the most insight into quantum algorithms, the quantum mechanisms that can be harnessed for computation, the breadth of applications of quantum computing, and the design of quantum processors.
Here, we describe efforts to provide partial answers to that question, focusing on those efforts which, through our involvement, we know best. 

We discuss novel gate-model approaches to exact and approximate optimization and sampling, particularly the quantum alternating operator ansatz. This extension of the quantum approximate optimization algorithm supports NISQ exploration of a broader array of problems, particularly those with hard 
constraints, by allowing more general mixing operators. This extension makes possible more efficient mixing that keeps the evolution within the feasible subspace, and supports mixing operators that are more easily implemented on NISQ hardware. We briefly touch on quantum annealing for network problems, for sampling with a quantum-assisted associative adversarial network, and for quantum variational autoencoders. Prior work on quantum annealing, including for planning and scheduling, fault diagnosis, and machine learning, was described in \cite{biswas2017nasa}.
We also describe novel classical and quantum methods for simulating many-body quantum systems, with potential applications to material science and chemistry, particularly  strongly-correlated quantum systems.
Quantum computing has inspired novel classical algorithms,
from optimization to the simulation of quantum systems. We focus on advanced quantum Monte Carlo (a quantum-mechanics-inspired classical algorithm) and other physics-based classical algorithms for optimization. These classical algorithms give insights into their quantum counterparts and also provide benchmarks against which any speedup claimed for a quantum algorithm can be assessed.  

Research into the roots of quantum computation in quantum physics and quantum information theory grounds the algorithmic work. We discuss recent studies providing a deeper understanding of thermalization in quantum annealers, making use of new features on the 2000Q D-Wave annealers
that support more flexible annealing schedules. We review recent work suggesting the potential to harness many-body delocalization effects for quantum optimization. 
Critical to this effort is the development of tools, including 
high-performance computing (HPC) simulation of quantum circuits, methods for compiling quantum circuits to realistic hardware, and state-of-the-art classical algorithms (including physics-inspired algorithms) against which to compare 
quantum algorithms.

This paper discusses contributions to: 
\begin{itemize}
\item Advancing gate-model quantum algorithms for
optimization, both exact and approximate, and for sampling
\item Advancing quantum annealing approaches to optimization and sampling
\item Physics-inspired classical algorithms for optimization and sampling
\item Quantum algorithms for simulating quantum many-body systems arising in material science and chemistry
\item Empirical testing of quantum approaches, both gate-model and annealing, applied to problems of interest
\item Novel methods, including temporal-planning-based methods, for compiling algorithms to near-term
quantum processors
\item Optimized HPC for simulation of near-term quantum circuits
\item Fundamental insights into the mechanisms of quantum computation.
\end{itemize}

The rest of this paper is organized as follows.
Sec.~\ref{sec:QAlgs} surveys recent work on quantum algorithms for optimization, both exact and approximate, and for sampling with applications to optimization and machine learning.
Sec.~\ref{sec:Qsim} turns to applications of quantum computing to simulating quantum systems, with applications in material science and chemistry.
Sec.~\ref{sec:Tools} covers tools to support quantum computing investigations going forward, including a novel HPC method for simulating quantum circuits, methods for compiling quantum circuits to near-term hardware, and quantum subroutines for scientific computing. 
Sec.~\ref{sec:mechQC} looks at recent work illuminating mechanisms of quantum computation, focusing on two topics, thermalization and many-body delocalization.
In Sec.~\ref{sec:future}, we conclude with some thoughts on the future outlook for quantum computing.

\section{Quantum Algorithms for Optimization and Sampling}
\label{sec:QAlgs}


At present, relatively few applications are known for which 
quantum computing provably outperforms classical (i.e. non-quantum) computation \cite{RPbook,nielsen2010quantum}.
Yet this scarcity of proven applications is unsurprising at this stage of the technology. 
Challenging computational problems arising in the practical world are
frequently tackled by heuristic algorithms. 
Heuristic algorithms often work well in practice, but by definition have not been analytically proven to be the best approach, or even proven to outperform the best previous approach on all instances.
Instead, heuristic algorithms are empirically tested on benchmark problems in, for example, satisfiability (SAT), planning, and machine learning competitions, and on real-world problems.
Due to the typically exponential slowdown for simulating quantum 
algorithms on classical hardware, prior to the emergence of quantum
hardware, quantum heuristics could only be tested on tiny problems,
usually capable of giving only inconclusive results.
Further, hardware constraints greatly limit the scope of  problems embeddable on existing quantum annealers, and it remains unclear to what degree these devices can provide advantages  
\cite{Ronnow2014defining,Boixo2014evidence,mandra2016strengths,denchev2016computational}.
Emerging universal quantum computers 
will enable a substantial broadening of the types of quantum heuristics that can be investigated. 
Empirical testing of quantum algorithms on quantum hardware in the coming decades will substantially broaden the applications for which quantum computing is known to outperform classical approaches. 
The immediate question is: what experiments should we prioritize that will give
us insight into quantum heuristics?


\subsection{The Quantum Alternating Operator Ansatz}

\label{sec:qaoa}


One leading candidate is the quantum approximate optimization algorithm (QAOA) metaheuristic. 
QAOA circuits were first proposed by Farhi \ea~\cite{Farhi2014}, 
subsequently leading to a number of tantalizing results%
~\cite{Wecker2016training,farhi2014quantum,farhi2016quantum,lin2016performance,Jiang17_Grover,guerreschi2017practical,Hadfield17_QAOAGen,verdon2017quantum,Wang18_QAOAFermionic}. 
At the high-level, QAOA circuits have a particularly simple form. 
Given a cost function $f(x)$ to optimize, 
a QAOA circuit alternates between
\lq\lq phase separation operators\rq\rq\ $U_P = e^{-i\beta H_P}$, which apply phases to computational basis states $\ket{x}$ depending on the cost $f(x)$,
and \lq\lq mixing operators\rq\rq\ $U_M = e^{-i\beta H_M}$, 
which generate transitions between the eigenstates of the cost-function Hamiltonian $H_P$.
QAOA circuits are parameterized by a set of real numbers 
$\{\boldsymbol{\gamma}, \boldsymbol{\beta}\}\equiv \{\gamma_j,\beta_j\}_{j=1}^p$, indicating the length of time the cost function Hamiltonian $H_P$ and the mixing Hamiltonian $H_M$ at each level are applied, 
where parameter $p$ is the number of iterations. 
A QAOA$_p$ circuit creates the quantum state $|\boldsymbol{\gamma \beta} \rangle
= U_M(\beta_p)U_P(\gamma_p)\dots U_P(\gamma_2)U_M(\beta_1)U_P(\gamma_1)|s\rangle$, 
where~$|s\rangle$ is a suitable initial state such as the uniform superposition state. 
By choosing a good parameter set,
quantum interference results in a concentration of amplitude in computational basis states with low cost.


Recent work 
\cite{Hadfield17_QAOAGen,Hadfield17_QApprox}
extends this framework to the 
Quantum Alternating Operator Ansatz (retaining the acronym QAOA), enabling alternation between more general families of operators. 
The essence of this extension is the use of alternating mixing and phase-separation operators drawn from more general one-parameter families of unitaries rather than only those corresponding to time-evolution under a fixed Hamiltonian for a time specified by the parameter. 
For example, $U_M(\beta)$ may take the form
$U_M(\beta) = U_m(\beta)U_{m-1}(\beta)\cdots U_1(\beta)$, where $U_i(\beta) = e^{-i\beta H_i}$,
where $\{H_i\}$, $1\leq i \leq m$, is a fixed set of Hamiltonians. 
This ansatz supports the representation of a larger, and potentially more useful, set of states than the original formulation, with potential for broad long-term impact 
on exact optimization, approximate optimization, and sampling. 
For many optimization problems, this extension allows simpler constructions, significantly expanding the implementability of QAOA on NISQ hardware.

Prior work on QAOA focused almost exclusively on cases in which all bit strings are feasible solutions,
and hence the mixing operator takes an exceptionally simple forms: independent flips of single qubits. 
When there are constraints that must be satisfied (hard constraints), 
it is  desirable to design the algorithm so that the state evolution stays within the feasible subspace, a subspace in which all constraints are satisfied, usually an exponentially smaller subspace of the Hilbert space (while still exponentially large itself). Our recent results on graph coloring problems confirm the advantages of mixers that constrain the evolution to the feasible subspace~\cite{wang19XY}. 
In~\cite{Hadfield17_QAOAGen}, we lay out design criteria for mixing operators.
We detail mappings for a diverse array of problems with different types of constraints and solution space structure,  including many prototypical NP-hard optimization problems such as Maximum Independent Set, Graph Coloring, Traveling Salesperson, and Single Machine Scheduling. 


Gate-model quantum computing is implemented through a sequence of local quantum unitaries, each involving a few qubits.
In QAOA, expanding the set of possible mixers
enables more efficiently implementable mixers, particularly for optimization problems with hard constraints. 
If a mixing Hamiltonian takes the form of a sum $H_M=\sum_k H_k$, 
where each $H_k$ involves a small number of qubits,
the Hamiltonian-based unitary $U_M = e^{-i\beta H_M}$ can still be quite
complicated to implement because
the $H_k$ typically do not mutually commute. 
A unitary of the form 
$U_M = \prod_k e^{-i\beta H_k}$ with any given ordering of $k$ also keeps the states in the feasible subspace so can serve as a mixing operator, but one that is often much easier to implement.


Current QAOA research  
includes mapping a wider variety of problems and applications, including problems
related to resource allocation and robust communication networks, and
exploring tradeoffs between alternative QAOA mappings. Evaluating the performance of QAOA on small problems using near-term hardware and classical simulation of quantum circuits, enables us to explore tradoffs between QAOA mappings with a variety of alternative mixers and initial states. Additionally, we continue to advance techniques for compiling QAOA circuits to NISQ hardware (see Sec.~\ref{sec:Compilation}). 

\label{sec:param_setting}

%
A critical research thrust is finding effective sets of parameters $\{\beta, \gamma\}$.
Generally, finding a global optimum $\{\beta,\gamma\}$ in parameter space 
is NP-hard (to see this, consider the circuit that rotates each
qubit individually into a bit string that minimizes the cost function).
However, a set of QAOA parameters does not have to be global-optimal 
in order for the resulting algorithm to demonstrate quantum advantage \cite{farhi2014quantum,Jiang17_Grover}.
Analysis of the algorithm and its parameter landscape benefits from identifying symmetries in the system \cite{Jiang17_Grover,Wang18_QAOAFermionic}. Understanding the landscape of parameter values of QAOA is a complicated quantum control problem, and numerical search can be hindered by not only local minima, but also by large flat areas in the energy landscape where the
gradient is very close to zero.
We illustrate these aspects in the following two examples.

\subsubsection{QAOA algorithm for Grover's search problem}

For a Boolean function~$f$ that takes value $0$ on all $n$ bit strings except for a single unknown bit string $x$, Grover's unstructured search problem \cite{grover1996fast} is to find the $x$ such that
$f(x)=1$, 
given only a black-box oracle for computing $f(x)$. 
Classical deterministic or randomized algorithms must access the oracle at least $\Theta(2^n)$ times to solve this problem in the worst-case, whereas Grover's algorithm \cite{grover1996fast,zalka_grovers_1999} showed a quantum computer requires only $\theta(2^{n/2})$ oracle calls. Thus, quantum computers achieve a quadratic improvement in query complexity for this problem. 
The QAOA approach~\cite{Jiang17_Grover} reproduces this quadratic improvement with an entirely new algorithm. 
It is also the first QAOA result for large $p < \infty$.
A unitary operator built from the standard mixing Hamiltonian $H_M$ and the cost (oracle) Hamiltonian $H_P$ are applied to the system periodically
\begin{align}\label{eq:W}
 W(\gamma) = e^{-i\pi H_M/n}e^{i\gamma H_P}e^{-i\pi H_M/n}e^{-i\gamma H_P}\,,
\end{align}
where there is only one free parameter $\gamma \in (0, \pi]$.
Analysis shows that 
such a quantum circuit yields a 
nearly optimal query complexity of $T\simeq \sqrt N (\pi/2\sqrt 2\,)$. Future work includes evaluating the effectiveness of periodic parameters for QAOA in other settings.

\subsubsection{Analysis of QAOA parameter landscape for the problem of ring-of-disagrees}
Another problem in which algorithm parameter setting and performance can be informed by
analytical treatment is 
maximum cut (MaxCut) on the $1$-d cycle (ring) graph,
known as the ``ring of disagrees'' \cite{farhi2014quantum}. 
On general graphs, MaxCut is NP-hard. 
For MaxCut on a ring, the optimal solutions are known: a bitstring of alternating values.
From numerical results for QAOA$_p$ on a ring with an even number of vertices~$n$, 
the approximation ratio achieved is $1$ for 
$p>n/2$,
and is $(2p+1)/(2p+2)$ for 
$p\leq n/2$  when the algorithm parameters are optimized~\cite{farhi2014quantum}. 
We apply a fermionic transformation to this problem in~\cite{Wang18_QAOAFermionic}, 
transforming it into 
a problem involving
$n/2$ rotating non-interacting individual spins.
All spins are initialized evenly on the unit circle in the x-z plane.
Applying the cost and mixing 
unitaries correspond to 
rotating each spin along the $z$-axis and a direction specific to each spin in the x-z plane, respectively.
The goal is to line up all spins along the $-z$-axis by the end of the algorithm.
From the symmetry observation that certain transformations bring the two axes on equal-footing,
we infer that optimal parameters must live in a submanifold defined by symmetry $\beta_i + \gamma_{p+1-i}=0$.
This reduces the number of parameters to search by half, and provides significant speedup in the search.

\begin{figure} [htbp]
	\begin{center}  
	\includegraphics[width=3in]{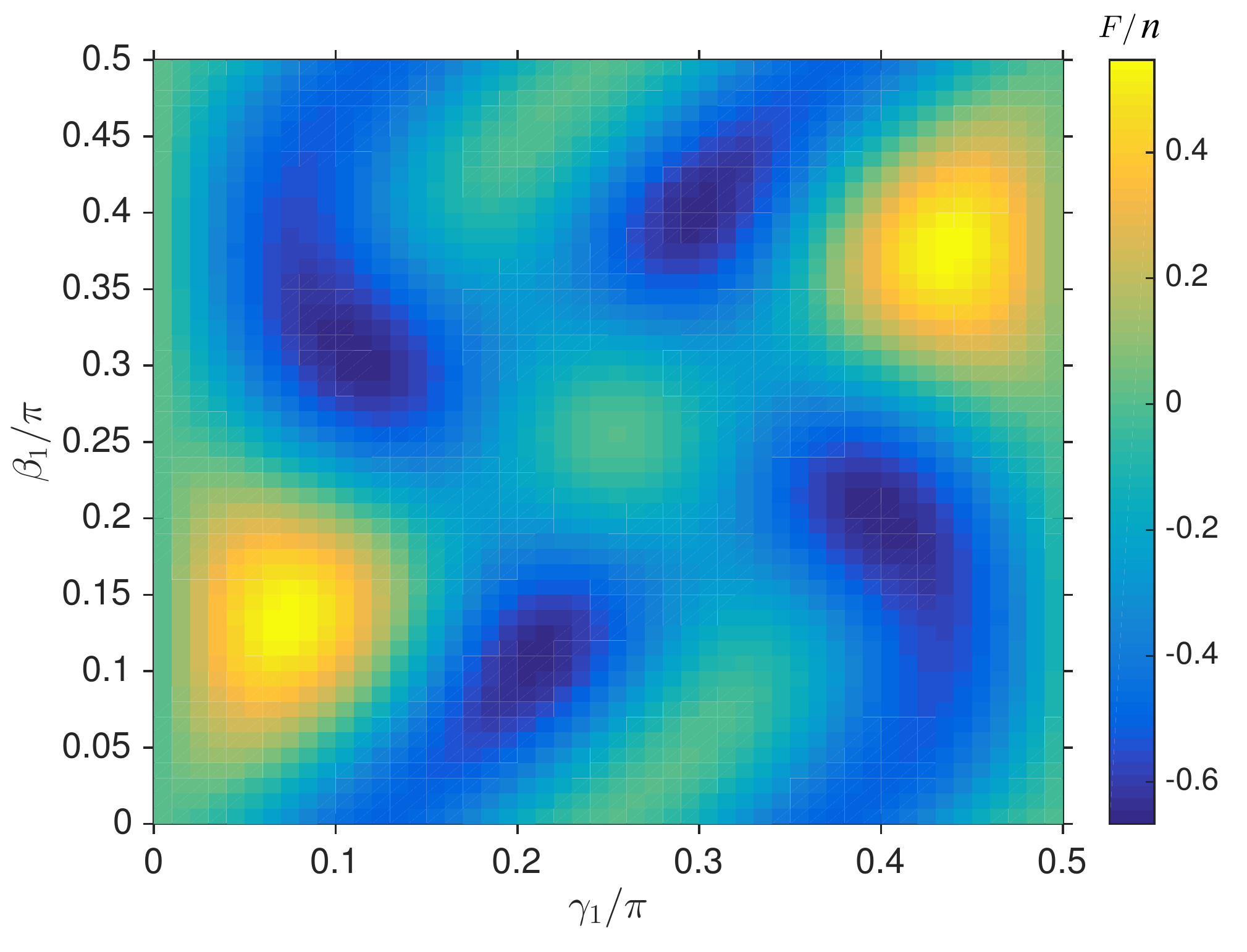}
	\caption{ (Color online).
Trap-free landscape of the space of parameter values for $QAOA_2$ on the problem of ring-of-disagrees, formulated as a minimization problem,  
in the sub-manifold $xx$. The four darkest spots indicate the global minima. The origin $(0,~0)$ is a saddle point.  No local minima are observed. 
}
\label{fig:p2as}
	\end{center} 
\end{figure} 


\subsubsection{Some general comments on parameter setting}
For parameterized quantum circuits, the landscape of the parameter space affects the ease with which good parameters can be found.  
We discuss two barriers that affect numerical search.
One is controllability and the existence of local minima,  and the other is barren plateau.
In quantum control theory, assuming
\emph{controllability} (i.e., evolution between any two states is achievable
via the set of controls given), the landscape of  parameter values generically has only global
minima~\cite{Rabitz04,Rabitz12,Rabitz16}.  Without controllability, the quantum
control landscape in general is rugged and admits local minima (\lq\lq traps\rq\rq)~\cite{Rabitz11}.
In the case of QAOA, the controls are constrained in a specific way: if
an infinite number of controls are allowed, i.e., $p\to \infty$, then the system
is controllable. The finite number of control steps dictated by finite $p$ limits the controllability.  
For QAOA on the problem of ring of disagrees, our results suggest that although the system is not controllable, within the symmetry-defined sub-manifold the landscape is trap-free, see~Figure~\ref{fig:p2as}. 
This simple problem demonstrates that the landscape of QAOA parameters can be complicated and calls for further studies 
into the relation with
quantum control theory.

Besides local minima, barren plateau may form another severe challenge.
In some circumstances, such as random quantum circuits, 
the gradient 
will be effectively zero over vast reaches of 
parameter space ~\cite{mcclean2018barren}, 
known as barren plateaus.
Barren plateaus foil gradient-based numerical 
optimization procedures. 
Although the framework of QAOA does not fall exactly into the conditions of this argument,
finding good parameters for QAOA may face a similar challenge.
Detailed study of quantum control is required to further understand both case-specific and 
general feature of the landscape of  parameter space for 
applications of QAOA, and as suggested in McClean et al.~\cite{mcclean2018barren}, good ans\"atze, such as those found in quantum chemistry, may be needed here. 

We are interested in analytical results and numerical 
experimentation that can lead to a better understanding of
how best to find good parameters for QAOA algorithms, design
trade-offs for the mixers, robustness under errors, and error mitigation
approaches tailored to these circuits. The ring of disagrees analysis
provides a promising start, as did the work of Yang et al.~\cite{Shabani16}.
We intend to explore further ideas for leveraging control theory, 
machine learning, and spectral properties and quantum transport techniques to provide deeper insights and better parameter-setting methods.

\subsection{Update on Quantum Annealing Work}
\label{sec:annealing}

Quantum annealing is a quantum metaheuristic for optimization~\cite{Nishimori98,farhi2000quantum}. 
Quantum annealers are quantum hardware that are designed to run this metaheuristic. 
Given a classical cost function $C(\bf x)$ with binary input ${\bf x} \in \{0,1\}^n$,
a quantum annealer works by encoding the binary variables $\bf x$ as qubits,
and slowly varying the Hamiltonian that governs the quantum evolution of the qubits
from a Hamiltonian $H_0$ whose ground state is the uniform superposition of all bit strings,
to a Hamiltonian $H_1$ that is constructed based on the cost function 
such that the global minimum of $C(\bf x)$ is the ground state of $H_1$. 

Quantum annealing has been a major research focus of the QuAIL team \cite{biswas2017nasa}. 
Our previous studies range from 
obtaining physical insights into and intuitions for quantum annealing,
studying programming and parameter setting for quantum annealers,
and application-specific as well as general classical solvers.
Older investigations of quantum annealing and quantum-classical hybrids to potential application areas, included 
planning and scheduling 
\cite{Stollenwerk17,Tran16a,Tran16b,Venturelli-JobShop-arXiv-2015,rieffel2015case,OGorman2015}, 
fault diagnosis \cite{Perdomo17_Readiness}, and machine learning \cite{Benedetti17,Perdomo17_Opportunities,Benedetti-2016,benedetti2017quantum,PerdomoOrtiz_EPJST2015}. In \cite{Jiang17_NonCommuting}, we pioneered the use of subsystem codes in the context of error suppression in quantum annealing, showing that subsystem codes circumvent a no-go theorem for stabilizer codes, enabling effective error suppression with just two-local terms. Marvian \cite{marvian2017error} has taken this approach further, deriving performance bounds and new two-local constructions.
We refer the reader
to \cite{biswas2017nasa} and references therein, for a survey of the older work.

Here, we look first at work on quantum annealing applied to robust network design, and then turn, in Sec.~\ref{sec:qa_ml} to a hybrid quantum-classical machine learning algorithm of the generative adversarial network (GAN) class. Later, Sec.~\ref{sec:mechQC} discusses lower-level advances related to quantum annealing that may be generally applicable.  
 
\subsubsection{Quantum Annealing for Network Problems}
\label{sec:network}


Driven by application-related network problems, 
we examined the spanning tree with bounded degree problem with quantum annealing.
As motivation, we start with a testing model in 
an aeronautic 
communication 
network problem.  
Consider a 
given number, say four, of unmanned vehicles with known trajectories and 
a pre-determined network disruption area $S$, 
where vehicles outside $S$ can talk directly to the ground control station 
but vehicles inside $S$ 
must route 
communication 
through vehicles not in $S$ to reach the ground station.  
If each vehicle can 
connect to at most $\Delta$ other vehicles (including the ground station), how 
should communications be routed 
so that the overall communication distance is minimized?  

This problem simplifies to minimizing the total edge weight of bounded-degree spanning trees of a graph specified by the problem instance. 
The spanning tree problem forms the basis of many network related applications. 
In the above case, the graph is a complete graph of five vertices (four vehicles and a ground station)
with some edges removed (between the ground station and the vehicles inside $S$) 
For 
an arbitrary graph, 
deciding whether or not a degree-constrained spanning tree exists is a NP-complete problem, and finding a degree-constrained spanning tree that optimizes certain 
cost functions 
on the graph is NP-hard.   
We developed three mappings of this problem to the quadratic unconstrained binary optimization (QUBO) form for
input to quantum annealer hardware.
These mappings can be readily embedded on the D-Wave annealer.  We performed sample runs for the different mappings 
on complete graphs of small size, and confirmed teh ability of the quantum annealer to solve these problems.

We describe one of these mappings, the ``level-based mapping,'' for a simple spanning-tree problem to illustrate the mapping process.
Given a graph $G=(V,E)$, $|V|=n,~|E|=m$, we aim to find a spanning-tree of $G$ through mapping it as a global minimum of QUBO.
Since a tree can be 
equivalently 
rooted at any node, we fix vertex $v=1$ to be the root.  We assign two types of binary variables.
For each vertex $v$, 
a binary variable $x_{u,v}$ for $\{u,v\}\in E$ indicates whether vertex $u$ is the parent of $v$,
and another binary variable $y_{v,l}$ indicates whether $v$ is at level $l$. 
We allow $l$ to run from $1$ to $n$.  The root $v=1$ is at level $1$ and has no parent.
Constraints that guarantee the graph to be a tree includes:
that each vertex $v\in(1,n]$ has exactly one parent; 
that each vertex $v$ is assigned exactly one level;
and that the parent vertex of $v$ must be of lower level than $v$.
The constraints are mapped to polynomial binary optimization terms as
$	(\sum_{v'} x_{v',v}-1)^2 $, 
$	(\sum_{l} y_{v,l}-1)^2  $, and
$	y_{v,l} \sum_{l'\ge l = 2}^n y_{v'l'} x_{v'v} \;,$
where the first two are in QUBO form and the last term can be further transformed into quadratic terms through ancilla qubits~\cite{rieffel2015case}.


\subsubsection{Quantum-Assisted Associative Adversarial Network Learning}
\label{sec:qa_ml}

Near-term utilization of quantum annealers in service of machine learning requires that the number of variables used be small and the algorithms resistant to noise. The aim of these algorithms is to turn over classically computationally expensive subroutines, such as sampling, to quantum hardware in hybrid architectures. Interest in integrating quantum Boltzmann machines and their restricted variants into deep learning architectures is driven by a potential increase in speed of sampling for some graph structures and the ability to learn quantum models.
We developed and evaluated a learning algorithm: Quantum Assisted Associative Adversarial Network (QAAAN) learning \cite{wilson2018}. This algorithm learns a latent variable generative model using a generative adversarial network (GAN) with an informed prior probability distribution.
In our approach, the canonical noise input to a generative adversarial network is replaced by samples from an informed prior. The prior is a model of a feature space taken from a low-dimension representation in the discriminative network and is learned by a Boltzmann machine. The output of the algorithm is a latent variable generative model, where sampling from the prior can be performed by sampling from the distribution with a quantum annealer or Markov chain Monte Carlo methods. We evaluated the algorithm on the MNIST, a standard machine learning dataset of images of handwritten digits. We found close performance between classical and quantum implementations when evaluated using the Frech{\'e}t inception distance (29 and 23, respectively) and the inception score (5.6 and 5.7, respectively), drawing the conclusion that the quantum device can be successfully integrated into the classical framework. We also demonstrated the classical architecture on the more complex dataset LSUN bedrooms, concluding the algorithm can be scaled. This work expands the set of deep learning algorithms that exploit near-term quantum annealers \cite{Perdomo17_Opportunities, khoshaman2018quantum, benedetti2018quantum}.


\subsubsection{Quantum Variational Autoencoders}
An Alternative Approach \cite{khoshaman2018quantum} to integrate quantum Boltzmann machines into deep generative models uses Variational Autoencoders (VAE) \cite{kingma2013auto} with discrete latent variables \cite{rolfe2016discrete, khoshaman2018gumbolt}. VAE are directed generative models that can be efficiently trained by maximizing a variational lower bound (ELBO) of the model log-likelihood. 
The use of variational inference and low-variance estimators of the training gradients \cite{kingma2013auto} makes this possible. In addition to achieving state-of-the-art performance in generative modeling (unsupervised learning), VAE are an important tool in semi-supervised learning \cite{kingma2014semi}.

When applied to natural image processing, VAE are very good at extracting coarse-grained features but struggle to reproduce the fine-details. This problem can be cured using a fully autoregressive decoder such as PixelCNN \cite{oord2016pixel,salimans2017pixelcnn++} at the cost of having a less informative latent space \cite{gulrajani2016pixelvae, chen2016variational}. We recently demonstrated that RBM priors enable an informative discrete latent space, state-of-the-art performance on common datasets such as MNIST and CIFAR10, and sharp image generation \cite{sadeghi2018}. This work demonstrates that quantum priors (which define distributions on discrete variables) can be used to train state-of-the-art quantum/classical hybrid generative models.

On important advantage of VAE is the potential to quantitatively evaluate trained models (by computing the ELBO or estimating the log-likelihood on the test set), enabling iterative improvement of the generative model. 
We  demonstrated \cite{vinci2019} using this approach that convolutional VAE can be successfully trained with D-Wave DW2000Q quantum annealers to achieve close to state-of-the-art performance on MNIST. The same paper demonstrates improved performance of VAE when using D-Wave if the structure of the convolutional encoder and decoder is crafted to the D-Wave Chimera architecture.

\section{Quantum Computing for Simulation}
\label{sec:Qsim}


The accurate simulation and characterization of systems of interacting fermions was one of the original motivations for the development of quantum computers~\cite{feynman1982simulating,Abrams1997,Ortiz2001}. 
There is optimism that for relatively near-term quantum computers that we will be able to perform some type of characterization of ground and excited state properties of quantum chemical systems~\cite{mcclean2016,babbush2018low-depth,peruzzo2014,kivlichan2018quantum,guzik2005}.  The field of interacting fermions (and other quantum particles) has applications in physics, chemistry and material science.
Due to broad applicability of fermionic simulations, efficient quantum simulation techniques would significant impact all of these fields. Recent estimates of the quantum resources \cite{Reiher2017} needed to perform phase estimation on FeMoco, a transition metal center that splits the nitrogen dimer in a nitrogenase enzyme, are  high compared to current quantum hardware, but represents a concrete goal for future quantum computers.

It is a challenge for quantum computers to achieve an advantage over classical computers
in this domain, given the decades of work on, and myriad classical tools for, treating quantum systems.  However, the quantum information processing viewpoint has proven fruitful for the discovery of advanced purely classical algorithms and analysis techniques (e.g.~\cite{drucker2009quantum,Kuperberg99,barak2015beating,tang2018Aquantum,tang2018Bquantum,gilyen2018quantum,vidal2004efficient}).  In this regard, our research focuses developing new techniques for both classical and quantum hardware from different perspectives.  This covers a range of systems including \textit{ab initio} simulations, Hubbard models, spin Hamiltonians, interacting Fermi liquids, solid state systems, and molecular systems~\cite{tubman2011,tubman2012,tubman2014,tubman2016,qmcpack,tubman2015,Tubman2018}.
While the long-term impact of quantum computing for simulations of quantum chemistry and materials is promising, we are in the early days; none of the experiments realized so far perform even
close to state-of-the-art classical simulations. For these reasons, near-term research will focus on advancing classical, and quantum-classical hybrid, as well as quantum algorithms, and on estimating the quantum resources, and designing quantum hardware architectures, for simulations of practical interest, so as to determine when and how such simulations can be realized.

An understanding of classical algorithms, and their strengths and weaknesses, can suggest where a quantum advantage may be most likely, as well as suggesting issues that may arise for quantum simulation approaches both in the near term and the long term.  In order to understand the feasibility of different quantum simulations broadly, we work at the interface between classical and quantum simulations. For example, in~\cite{Tubman2018} we used leading classical simulations to start providing resource estimates for aspects of quantum simulations that are crucially important but often neglected: 
initiating the starting wave function used in quantum methods such as phase estimation.  
Efficient \textit{state preparation}, finding and creating an effective initial state, must be part of any efficient quantum simulation algorithm. 
We designed
a new approach to creating correlated wave functions on a quantum device, 
and demonstrated this  new approach to generating effective initial states with classical algorithms.    
We investigated state preparation for a series of difficult physics- and chemistry-relevant Hamiltonians, demonstrating the plausibility of generating good initial states for problems that may not be solved to high accuracy with current classical algorithms.  

Other recent work \cite{hadfield2018divide} showed that faster quantum simulation algorithms may be obtained by considering Hamiltonian structure. For simulating the sum of many Hamiltonians, we split the Hamiltonians into groups, simulate each group separately, and then combine the partial results, with each simulation tailored to the properties of each group. 
We illustrated our results using the electronic structure problem of quantum chemistry, in which the second-quantized Hamiltonian is typically given by a sum in which relatively few terms are substantial in norm. Separately grouping the large norm and small norm terms significantly improved simulation cost estimates under mild assumptions.

Another promising research direction is improving existing quantum methods 
Two of the most discussed methods, phase estimation and variational quantum eigensolvers (VQE) require significant quantum computing resources, prompting work to find less resource-intensive alternatives.  
For example, while the variational quantum eigensolver~\cite{mcclean2016,peruzzo2014,yung2014transistor} approach has 
early experimental demonstrations~\cite{peruzzo2014,OMalley2016,Santagati2018,Colless2018,kandala2017}, it is far from settled whether or not it will be possible to overcome  difficulties involved in choosing a useful ansatz, optimizing many thousands of parameters~\cite{mcclean2018barren}, and scaling of noise and measurement challenges \cite{wecker2015progress,Rubin2018,mcclean2017hybrid}. 
Some of these challenges were addressed for spinless Hubbard models, where a suitable circuit's structure was proposed based on the entanglement properties of the examined Hamiltonians and achievable speed of convergence~\cite{Woitzik2018, Woitzik2019}.
Much research remains to be done in this direction, with many wave function ans\"atze to explore~\cite{huggins2018} to advance  variational quantum eigensolvers.

Phase estimation, a method for determining eigenstate energies, is useful for 
the accurate simulation of long-time dynamics and the determination~\cite{Kitaev1995,Abrams1999}, but requires a large number physical qubits and quantum gates~\cite{Jones2012,wecker2014,Reiher2017,BabbushSpectra}. 
While quantum phase estimation has been experimentally implemented for small molecules on a variety of quantum computing architectures, including with quantum optics \cite{lanyon2010}, nitrogen-vacancy centers \cite{Wang2014} and superconducting qubits \cite{OMalley2016}, its application to larger molecules of practical interest, such as FeMoco, is estimated to need a gate depth several orders of magnitude beyond what is currently feasible~\cite{Reiher2017}.  We are therefore interested in trying to improve techniques for these types of simulations.  

In addition to finding new algorithms for quantum simulation, tools for maximally utilizing NISQ hardware, such as methods for compiling quantum circuits to near-term architectures, can reduce the quantum resources required compared to more straightforward implementations. 
Sec.~\ref{sec:JWcompilation} describes one novel approach to compilation of circuits for simulation (and also for other near-term algorithms such as QAOA). 

\section{Tools for Quantum Computing Investigations}
\label{sec:Tools}

Quantum computing research requires an array of tools, and the creation of new or improved tools. Here, we discuss advances in methods for simulating quantum circuits on classical computers, in physics-inspired optimization algorithms for benchmarking quantum algorithms, in compiling quantum circuits to near-term hardware, and in efficient implementation of subroutines commonly used in scientific computing.

\subsection{Classical Methods for Simulating Quantum Circuits}

Classical simulation of quantum circuits not only provides insight into quantum heuristics: it is key to validating
early quantum hardware 
\cite{boixo2018characterizing,Neill17}.
The direct comparison of results from NISQ devices and from classical simulations will enable prioritization of hardware improvements 
to best support these algorithms. The computational effort involved in simulating quantum circuits grows exponentially with the number of qubits, making even low-depth circuits challenging to simulate
\cite{boixo2018characterizing,harrow2017quantum,bravyi2018quantum}. 
In recent years, new classical techniques have been devised to
take advantage of regular structures of near-term implementable quantum algorithms 
such as QAOA, instantaneous quantum computation (IQP), 
random circuits, and Variational Quantum Eigensolvers (VQE). Among the classical techniques
to simulate quantum circuits, the most promising ones to simulate near-term
devices are the tensor network contraction \cite{markov2008simulating} and 
the Bravyi-Gosset stabilizer approach \cite{bravyi2016improved}. While the
former technique is more suitable for shallow quantum circuits, the latter
works the best for deep quantum circuits with non-Clifford gates.\\

\begin{table*}
\centering
\begin{tabular}{|c|c|cc|cc|}
\hline 
\multirow{2}{*}{\textbf{Circuit size}} & \multirow{2}{*}{\textbf{Target fidelity (\%)}} & \multicolumn{2}{c|}{\textbf{Runtime (hours)}} & \multicolumn{2}{c|}{\textbf{Energy cost (MWh)}}\tabularnewline
 &  & \textbf{Pleiades}  & \textbf{Electra }  & \textbf{Pleiades }  & \textbf{Electra}\tabularnewline
\hline 
\hline 
$7\times7\times(1+40+1)$  & 0.51 & 62.4  & 59.0  & $2.73\times10^{2}$  & 96.8 \tabularnewline
$8\times8\times(1+32+1)$  & 0.78 & 1.38  & 1.59  & 6.91 & 2.61 \tabularnewline
$8\times8\times(1+40+1)$  & 0.58 & $1.18\times10^{4}$  & $1.23\times10^{4}$  & $5.15\times10^{4}$  & $2.02\times10^{4}$\tabularnewline
$8\times9\times(1+32+1)$  & 0.51 & 14.8  & 15.2  & 73.9  & 24.9 \tabularnewline
Bris.-70 $\times(1+32+1)$  & 0.50 & 145  & 178  & 723  & 293\tabularnewline
\hline 
\end{tabular}
\vspace{4pt}
\caption{\label{table:classical}Estimated runtimes and energy costs for the
computation of $10^6$ amplitudes with fidelity close to $0.5\%$ on NASA HPC
Pleiades and Electra systems. The $7\times 7\times (1+40+1)$ and
$8\times 8\times (1+40+1)$ grid jobs do not fit in Sandy Bridge nodes, due to
their memory requirements; for that reason, the portion of Pleiades with Sandy
Bridge nodes is not considered, in the energy cost estimations for these two cases.}
\end{table*}

\begin{table*}
\centering
\begin{tabular}{|c|c|cc|cc|}
\hline 
\multirow{2}{*}{\textbf{Circuit size }} & \multirow{2}{*}{\textbf{Fidelity (\%) }} & \multicolumn{2}{c|}{\textbf{Runtime (hours)}} & \multicolumn{2}{c|}{\textbf{Energy cost (MWh)}}\tabularnewline
 &  & \textbf{Pleiades } & \textbf{Electra } & \textbf{Pleiades } & \textbf{Electra }\tabularnewline
\hline 
\hline 
$7\times7\times(1+40+1)$  & 100.0  & $1.22\times10^{-2}$  & $1.16\times10^{-2}$  & $5.35\times10^{-2}$  & $1.89\times10^{-2}$ \tabularnewline
$8\times8\times(1+32+1)$  & 100.0  & $1.77\times10^{-4}$ & $2.04\times10^{-4}$  & $8.86\times10^{-4}$  & $3.34\times10^{-4}$ \tabularnewline
$8\times8\times(1+40+1)$  & 100.0  & $2.03$  & $2.12$  & $8.88$  & $3.48$\tabularnewline
$8\times9\times(1+32+1)$  & 100.0  & $2.90\times10^{-3}$  & $2.98\times10^{-3}$  & $1.45\times10^{-2}$  & $4.89\times10^{-3}$ \tabularnewline
Bris.-70 $\times(1+32+1)$ & 100.0  & $2.89\times10^{-2}$  & $3.57\times10^{-2}$  & $1.45\times10^{-1}$  & $5.85\times10^{-2}$ \tabularnewline
\hline 
\end{tabular}
\vspace{4pt}
\caption{\label{table:verification} Estimated effective runtimes and energy cost
for the computation of a single amplitude with perfect fidelity on NASA HPC
Pleiades and Electra systems. The $7\times 7\times (1+40+1)$ and
$8\times 8\times (1+40+1)$ grid jobs do not fit in Sandy Bridge nodes, due to
their memory requirements; for that reason, the portion of Pleiades with Sandy
Bridge nodes is not considered in the energy cost estimations for these two cases.}
\end{table*}

Utilizing NASA Ames's supercomputer resources and expertise to optimize code based on the aforementioned techniques, in collaboration with the Quantum AI team
at Google, and ORNL\cite{villalonga2018, villalonga2019}, 
we devised a simulator that pushes the boundary
on the size and complexity of near-term quantum circuits that can be 
simulated. The resulting simulator can simulate random circuits on Google Bristlecone QPU 
\cite{boixo2018characterizing,GoogleBristleconePreview}. 
The simulator can compute both exact amplitudes, a task essential for the verification of the
quantum hardware, as well as low-fidelity amplitudes to mimic NISQ devices. For instance,
our simulator is able to output single amplitudes with depth 1+32+1 for the full $72$-qubit
Google Bristlecone QPU in less than $(f \cdot 4200)$ hours on a single core,
where \mbox{$0 < f \leq 1$} is the target fidelity, on $2\times20$-core Intel Xeon Gold
6148 processors (Skylake). We also estimate that computing $10^6$ amplitudes
(with fidelity 0.50\%) needed to sample from the full Google Bristlecone QPU
with depth (1+32+1) would require about 3.5 days using the NASA Pleiades and
Electra supercomputers combined. 
Table~\ref{table:classical} and Table~\ref{table:verification} report
our latest numerical estimations \cite{villalonga2018} to classically simulate
quantum random circuits for 2-D grids, including the Google Bristlecone QPU.
Tables I and II report estimated runtimes and energy costs for 
the computation of $10^6$ noisy amplitudes, with target fidelity of $0.5\%$ (Table I)
or single perfect amplitudes (Table II). As comparison, the power needed by 
1 million inhabitants city is $\sim$1500 MW.
Our analysis is supported
by extensive simulations on NASA HPC clusters Pleiades and Electra
(further details in \cite{villalonga2018}).
For the most computationally demanding simulation we had, the two HPC clusters combined reached a
peak of 20 Pflop/s (single precision), that is $64\%$ of their maximum achievable
performance. To date, this numerical computation is the largest in terms of
sustained Pflop/s and number of nodes utilized ever run on NASA HPC clusters.
We also performed numerical benchmarks on Summit, currently the fastest supercomputer in the
world\cite{villalonga2019}. After porting qFlex for GPU architectures, we were able to reach
an average performance of 281 Pflop/s (true single precision, corresponding to the 68\%
of the maximum achievable), with peaks of 381 Pflop/s (true single precision, corresponding
to the 92\% of the maximum achievable).
All numerical simulations can be found at the QuAIL@NASA repository~\cite{note:db_release}.\\

State of the art algorithms for simulating quantum algorithms are \emph{parameterized}; their resource requirements scale exponentially only in some property of an instance rather than its size, e.g. the number of T-gates for the stabilizer approach.
Tensor network contraction was shown to be parameterizable by the treewidth of the line graph of the tensor network's graph; we found an alternative characterization of this result in terms of tree embeddings and vertex congestion\cite{ogorman2019parameterization}.
The characterization in terms of vertex congestion is conceptually clearer and exactly quantifies the running time when the contraction is done as a series of matrix multiplications. 
It also yields a charaterization, in terms of edge congestion, of the memory used, which is often the bottleneck in practice.

\subsection{Physics-inspired Classical Algorithms and Benchmarking of
Quantum Algorithms and Applications}
\label{sec:classicalAlgs}


We continue to push the state-of-the-art in classical optimization
techniques, particularly physics-inspired techniques. 
Quantum Monte Carlo 
\cite{Jiang17_Instanton, Jiang17_PathIntegral, kechedzhi2018efficient}
and iso-energetic cluster moves 
\cite{zhu2015efficient}, 
provide a fair comparison of quantum heuristics to classical techniques
\cite{mandra2016strengths, Mandra17_Pitfalls, mandra2018deceptive}
and a reference for claims of quantum advantage.
Key to evaluation of quantum advantage is characterizing different kinds
of quantum speed-ups \cite{mandra2016strengths,Ronnow2014defining}.

In addition to both classical and quantum heuristics for hard-optimization
problems, we developed novel quantum and hybrid quantum-classical 
approaches for real-world applications in the areas of planning and scheduling
\cite{Stollenwerk17,Tran16a,Tran16b,Venturelli-JobShop-arXiv-2015,rieffel2015case,OGorman2015}, 
fault diagnosis \cite{Perdomo17_Readiness}, and machine
learning applications \cite{benedetti2017quantum, Benedetti-2016, Perdomo17_Opportunities}.
In contexts in which diverse solutions are valued, aspects of the cost function are not available ahead of tiem, or the problems change quickly  with time,
having multiple optimal and quasi-optimal solutions is of utmost importance.
To this end, we are also advancing classical optimization approaches
through physics-based advances in classical algorithms for sampling
\cite{Mandra17_Sampling,ochoa2018feeding}.

\begin{figure}
    \includegraphics[width=0.5\textwidth]{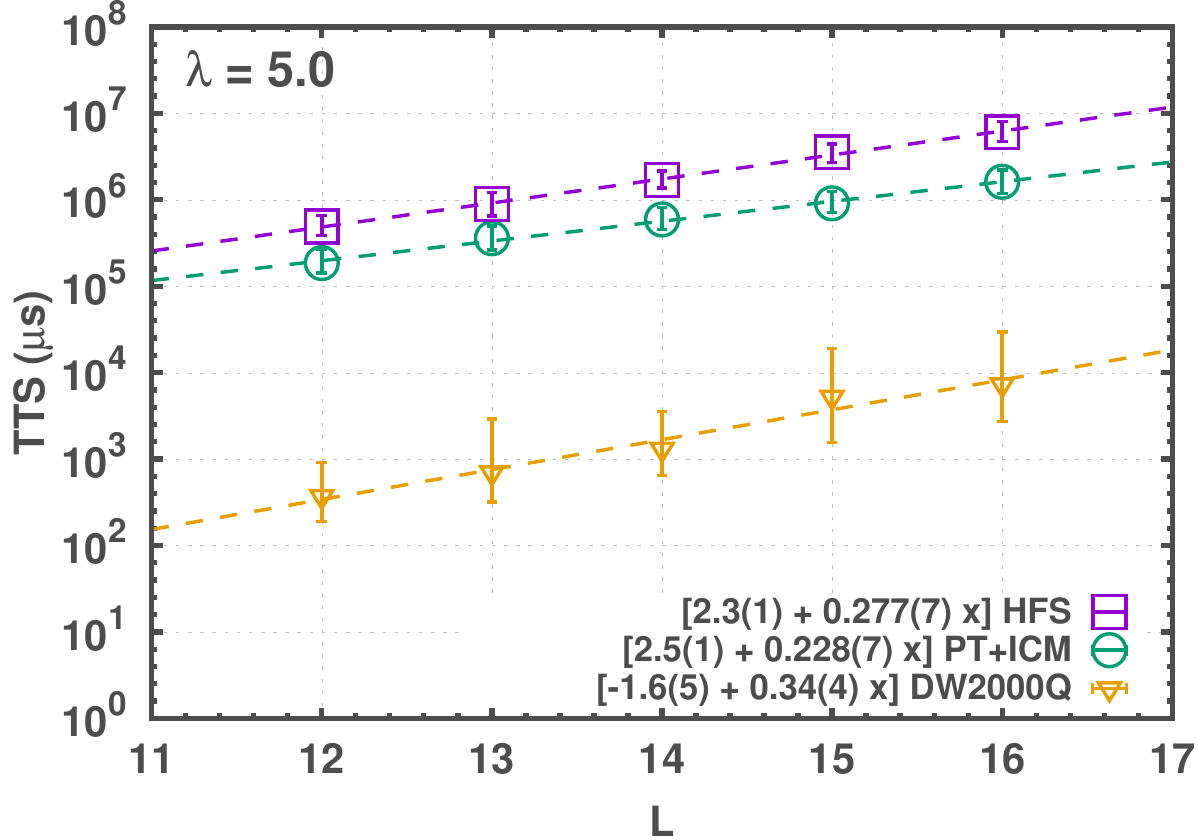}
    \caption{\label{fig:deceptive}Time-to-solution (TTS) for the parallel tempering iso-energetic 
    cluster method (PT+ICM), the Hamze-de Freitas-Selby (HFS) heuristic, as well for the D-Wave 2000Q
    (DW2000Q) quantum chip (see \cite{mandra2018deceptive} for more details).}
\end{figure}
As part of the effort to detect and identify quantum speed-ups, 
the QuAIL team's work developing benchmark problems sets including
the theory, identification, and generation of parameterized families of hard 
problems \cite{mandra2018deceptive}, containing small problems suitable for near-term, 
prototype quantum processors \cite{Rieffel14,Wang17_SMSPT}, generation of
problems with planted solutions and fine-tuned complexity 
\cite{Wang17_Patch}. Figure~\ref{fig:deceptive} shows an example
in which the D-Wave quantum device shows a better pre-factor (but a similar
scaling and comparable energy consumption) if compared to the 
state-of-the-art classical heuristics once the problem parameters are properly tuned.
We will use these benchmark problem sets for evaluating early quantum
hardware, both to evaluate the utility of annealing capabilities newly available on the D-Wave quantum annealers
such as the anneal offset, pause and quench, and reverse
annealing \cite{marshall2018power}, and to evaluate the effectiveness of gate-model approaches
such as QAOA (see Sec. \ref{sec:qaoa}).
We will extend this approach to the evaluation of
quantum-classical hybrid methods as well. 

\subsection{Compiling Quantum Circuits to Realistic Hardware}
\label{sec:Compilation}

Compilation is required to take descriptions of quantum algorithms and generate implementations on near-term hardware. There are multiple aspects of compiling, including addressing the limited connectivity of many NISQ devices. One way to overcome limited connectivity is to insert swaps in the circuit so that states can be moved to adjacent qubits where logical operations can be performed. The problem of doing so optimally is referred to as the “qubit routing” problem.Here, we consider two types of routing approaches, one based on temporal planning, the other on Jordan-Wigner reordering.

\subsubsection{Temporal Planning for Quantum Circuit Compilation}
\label{sec:TPcompilation}

Temporal planning is an effective approach for qubit routing of
quantum circuits to near-term hardware  
\cite{Venturelli18_Compiling}.
Temporal planning ~\cite{ghallab2004automated}, a subdomain of Automated Planning and Scheduling, 
is concerned with choosing actions,
among a large finite set of possibilities, to achieve a specific goal, incorporating temporal constraints or time optimization objectives. 
Our mapping of compilation to planning allows us to exploit 
highly optimized, off-the-shelf planners.
Temporal planners minimize the 
{\it makespan}, a refinement
of circuit depth that takes into account gate times, namely the time it takes to run a circuit. NISQ hardware does not
have the resources to implement error correction, and thus suffers from decoherence, making the makespan a quantity of critical relevance.

The temporal-planning approach is particularly effective for 
circuits, such as QAOA circuits, with many commuting gates, in which 
the space of possible compilations is larger than for circuits with more restricted ordering. 
More recently, we integrated the temporal
planning approach with constraint programming \cite{Booth18}, 
and have started working on more sophisticated hybrids of temporal planning
and constraint programming (CP), including adaptive switching approaches. Results are encouraging, as temporal planning if properly tuned and
hybridized with CP, can deliver provably optimal results
for the instances tested in~\cite{Venturelli18_Compiling}.

\begin{figure}[t]
    \includegraphics[width=\textwidth]{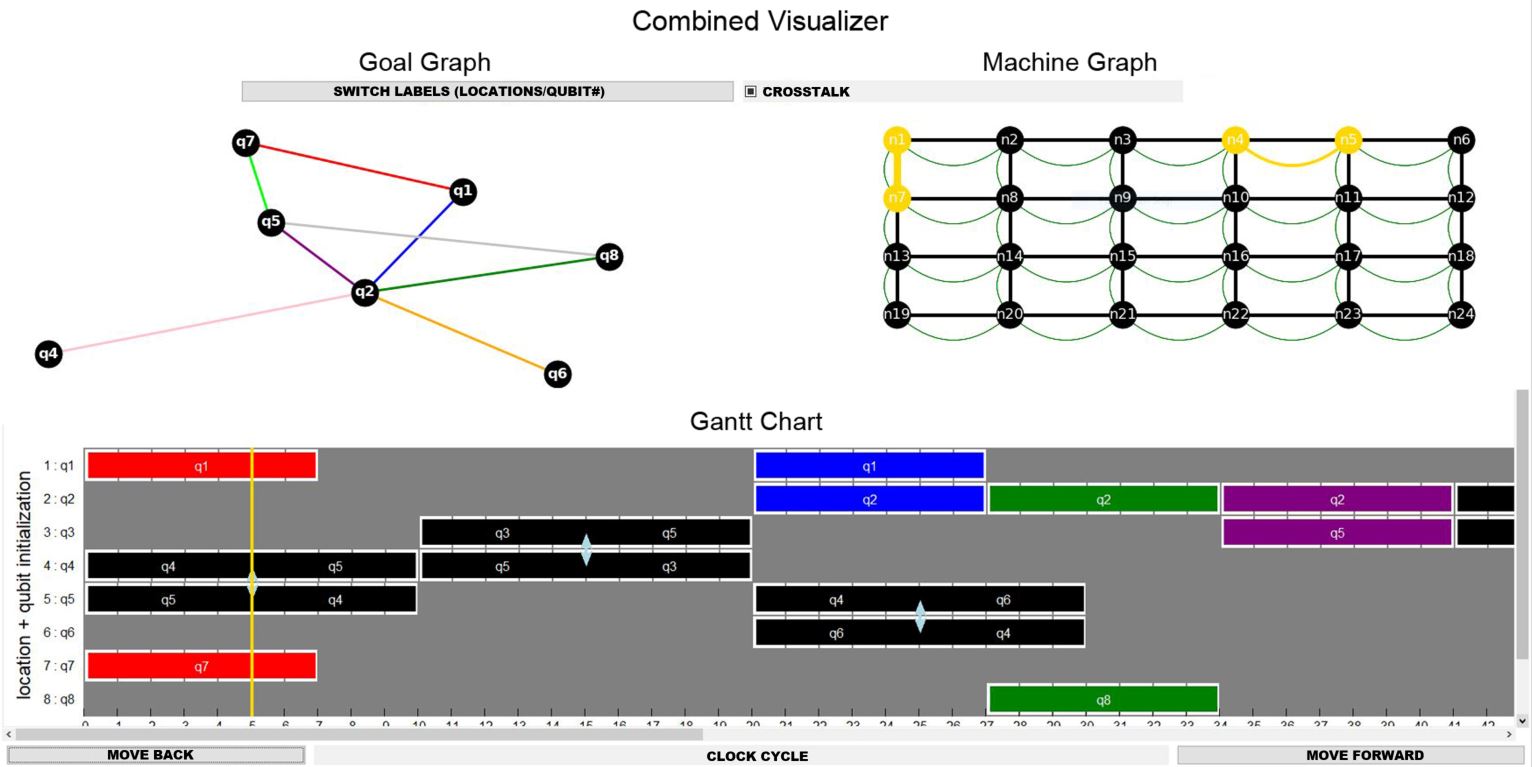}
    \caption{\label{fig:compiler}Snapshot of the compiler visualizer interface. Top left shows the input graph (MaxCut application). Top right visualizes
    a Bristlecone 24-qubit architecture, at a point in time where two 2-qubit gates are active. Bottom: Gantt chart of the compiled circuit (see~\cite{Venturelli18_Compiling} ).
    Each time can be selected and inspected.}
\end{figure}

We developed software components for a fully-fledged module that supports temporal-planning and CP-based compilation to NISQ processors, currently instantiated for Google's Bristlecone architecture and MaxCut problems. 
The software supports visualization (Fig.~\ref{fig:compiler}) and interactive tuning of the compiled circuits. See \cite{SPARK2019} for details.
We are extending our
evaluation from QAOA MAXCUT circuits to QAOA circuits for a variety of problems
\cite{Hadfield17_QApprox,Hadfield17_QAOAGen}, starting from Graph Coloring.
The automated reasoning framework is broad enough to incorporate more hardware
constraints and capabilities such cross-talk, measurement, and feedback.
Another important research direction is to extend this approach beyond superconducting architectures.


\subsubsection{Jordan-Wigner-Based Compilation for Quantum Simulation and Beyond}
\label{sec:JWcompilation}

The near-term algorithms in Sec.~\ref{sec:QAlgs} and Sec.~\ref{sec:Qsim} are typically based on one of two primitive steps: Trotterized time evolution or Hamiltonian-based variational ans\"atze.
From a compilation perspective, these have essentially the same structure, namely a gate for each term in the Hamiltonian.
While these gates do not necessarily commute, there is significant freedom in how they are ordered, and compilation efficiency may be the dominant criterion therefor.
Using the standard Jordan-Wigner transformation from fermionic operators to qubit operators, each term in the Hamiltonian corresponds to a gate whose locality depends on the differences of the indices in the relevant fermionic modes.
However, fermionic swap gates can be used to reorder the indexing of the modes so that each gate has minimum locality (i.e.\ it acts on the same number of qubits as the corresponding number of fermionic modes).
Previous work~\cite{babbush2018low-depth} showed how to compile circuits corresponding to quadratic fermionic Hamiltonians on $n$ modes in $O(n)$ depth.
However, these quadratic Hamiltonians derive from particular bases that are likely unsuitable for molecular systems.
We generalized the approach to arbitrary bases, showing how to get the optimal depth of $O(n^3)$ for a general chemical Hamiltonian \cite{ogorman2019generalized}.
Using similar methods, we also show how to achieve $O(n^2\eta)$ depth for Unitary Coupled Cluster with $n$ modes and $\eta$ electrons, as well as $O(n)$ depth for Unitary Paired Coupled Cluster \cite{huggins2018}.
Importantly, the optimality of this scaling is independent of the connectivity of the hardware and shows that the parallelization can address the locality problem of the Jordan-Wigner transformation for free.
For ``low-rank'' instances, other approaches~\cite{motta2018low-rank} may provide more efficient compilations, but our techniques are unconditional.

This approach to compiling fermionic gates by reordering the Jordan-Wigner string to get certain modes adjacent~\cite{kivlichan2018quantum} is equivalent to compiling local qubit operators to a physical line of qubits.
Therefore the techniques are equally applicable, e.g., to QAOA circuits for constraint satisfaction problems.
In this context, the approach can be considered an instance-independent compilation; for any particular instance, the compilation may be suboptimal, but with the advantage that there is no instance-specific compilation time, which is important for practically benchmarking against classical methods.

\subsection{Quantum Subroutines for Scientific Computing}


Algorithms for scientific computing applications often require modules, i.e., building blocks, implementing elementary numerical functions that have well-controlled numerical error, are uniformly scalable and reversible, and that can be implemented efficiently. 
We showed explicit quantum circuits, at the level of basic addition and multiplication operators,
for a variety of numerical functions important in 
scientific computing in \cite{hadfield2016scientific}. The
decomposition into basic addition and multiplication operations, and hence the
corresponding error and cost bounds, are largely agnostic to the underlying
hardware. This allows for easy portability between different quantum
architectures, and straightforward derivation of the cost in terms of primitive
quantum gates. 

In particular, in  \cite{hadfield2016scientific} we
derive quantum algorithms and circuits for computing square roots, logarithms,
and arbitrary fractional powers of numbers, and derive worst-case error and cost bounds.
We describe a modular approach to quantum algorithm design as a first step
towards numerical standards and mathematical libraries for quantum scientific
computing. 
Ongoing work seeks to explore the
tradeoffs between different low-level realization of these circuits, such as between circuit depth and the number of qubits required for scratchpad space, and between the resources required for different implementations of the underlying quantum circuits for addition and multiplication.

In \cite{hadfield2018boolean}, we showed a methodical approach to mapping
classical Boolean and real functions to quantum Hamiltonians in terms of Pauli
operators. Quantum circuits for simulating these Hamiltonians and their costs
then follow immediately from these mappings. In particular, this allows for
efficient construction and of a wide variety of QAOA phase operators. Nevertheless, 
it remains an ongoing research direction to explore when these constructions are optimal with respect to a fixed gate-set, i.e., 
when the cost of the construction obtained from the Hamiltonian mapping cannot be improved upon significantly. 
Furthermore, the results of \cite{hadfield2018boolean} are also applicable to quantum annealing, where they give a methodical approach to mapping general objective functions to Hamiltonians acting on quantum spins, and to constructing Hamiltonians implementing penalty functions in the case of constrained optimization.  

\section{Mechanisms of Quantum Computation}
\label{sec:mechQC}

Here, we discuss recent work that deepen understanding and quantum mechanisms. The first is research into thermalization in quantum annealers and the effect of annealing schedules on performance. The second is many-body delocalization with potential implications for quantum computing.
An overview of our past work, including \cite{mandra2016faster,Marshall17,Mandra17_Sampling,Smelyanskiy17_Environment,
Jiang17_Instanton,Jiang17_PathIntegral,Kechedzhi16,Venturelli15,Boixo2014evidence,Smelyanskiy18_NonErgodic} can be found in \cite{biswas2017nasa}.

\subsection{Thermalization and Sampling}
\label{sec:thermal}


In \cite{marshall2018power}, we investigated alternative annealing schedules on current generation 2000 qubit quantum annealing hardware (the D-Wave 2000Q), including the use of forward and reverse annealing with an intermediate pause. 
This work provides new insights into the inner workings of these devices, and quantum devices in general, particularly into how thermal effects govern the system dynamics. 
On benchmark problems native to the D-Wave architecture, we showed that a pause mid-way through the anneal dramatically changes the output distribution. Upon pausing the system in a narrow region shortly after the minimum gap, the probability of finding the ground state of the problem Hamiltonian can be increased by several orders of magnitude,  compared to not pausing or when pausing outside of this region. By identifying three relevant time-scales, we provide evidence suggesting thermalization is indeed occurring during such a pause. 
We related this effect to relaxation (i.e. thermalization) after diabatic and thermal excitations that occur in the region near the minimum gap. For a set of small native problems, for which the minimum gap can be computed, we confirmed this relation. For a set of large-scale native problems of up to 500 qubits, we demonstrate that the distribution returned from the annealer closely matches a (classical) Boltzmann distribution of the problem Hamiltonian, albeit one with a temperature higher than the (effective) temperature of the annealer.
An encouraging result is that larger problems appear more likely than smaller problems to thermalize to a classical Boltzmann distribution, suggesting the potential use of quantum annealers as Boltzmann sampling subprocessors in quantum-classical hybrid machine learning algorithms.
Our model is general, suggesting that similar thermal effects could apply to different types of quantum devices, and within quantum annealing, the results should hold for other classes of problem.
This work suggests that not only pausing, but alternative annealing schedules more generally, are a promising way to improve time-to-solution for a wide range of optimization problems.

\subsection{Many-body Delocalization}


A deeper understanding of many body dynamics will aid quantum computing, from hardware design to novel algorithms. Of particular interest are quantum effects that could be harnessed for quantum computational purposes. Many-body delocalization is potentially one of those effects.
Many-body localization (MBL) is the robust breakdown of ergodicity in quantum disordered systems, when an isolated quantum system does not properly function as a bath for its own local
subsystems, which therefore fail to thermalize \cite{d2016quantum}.
Localization/delocalization transitions are typically driven by a parameter controlling
the relative strength of kinetic and disorder terms affecting a quantum system. In
the strong disorder phase (also called the localized phase), these systems develop quasi-local
integrals of motion \cite{huse2014phenomenology,serbyn2013local,ros2015integrals} that
constrain its dynamics in a ``frozen'' regime where transport is suppressed. This is
detected by localized spatial configuration of the eigenfunctions of the system's Hamiltonian:
delocalized systems have wavefunctions whose amplitudes are roughly equal 
($| \psi_i |^2 \approx 1/\mathrm{Vol}$)
over all spatial sites, while the amplitudes of localized
wavefunctions concentrate in a finite region and  decay exponentially 
away from it. Although originally defined for tight-binding models of electrons hopping on a
lattice \cite{basko2006metal}, MBL was 
generalized to quantum spin models and spin
glasses \cite{imbrie2014many,imbrie_diagonalization_2016}.

The connection between localization and quantum computing was first drawn in \cite{Altshuler12446}. 
Writing the standard quantum annealing Hamiltonian $H(t) = H_{cl} + B_{\bot}(t)\sum_i \sigma_i^x$ in the (classical) computational basis, one recovers the Hamiltonian
\begin{equation}\label{eq:Bool_hpcube}
H = \sum_{i}\epsilon_i | i \rangle \langle i | + B(t) \sum_{\langle i,j\rangle} \Big(| i \rangle \langle j| + | j \rangle \langle i | \Big),
\end{equation}
where the sum is over all pairs $\langle i,j \rangle$ of classical states connected by exactly one spin flip (a \emph{Boolean hypercube} graph structure).
The Hamiltonian of Eq. (\ref{eq:Bool_hpcube}) is analogous to the Anderson model Hamiltonian
\begin{equation}\label{eq:Anderson_model}
H_{AM} = \sum_i \epsilon_i a_i^{\dagger}a_i + t \sum_{\langle i,j\rangle} \Big( a_i^{\dagger}a_j + a_j^{\dagger}a_i \Big)
\end{equation}
restricted to the one-particle sector of its Fock space. The Anderson model describes the theory of free electrons hopping in a $d$-dimensional cubic lattice, scattered by a disordered local potential with uncorrelated energies $\epsilon_i$. It undergoes a localization/delocalization transition in dimensions $d \geq 3$. An open question is whether the same is true for the 
when the energies $\epsilon_i$, the costs that appear in the (local) combinatorial optimization problem $H_{cl}$, show correlations. The emergence of localized dynamics for such systems would suppress quantum tunnelling and thus negatively impact the performance of quantum annealers in the part of the anneal when $B(t)$ is small.

In light of the similarity between the spectra of various spin glasses models and combinatorial optimization problems, the current effort by the scientific community to detect the appearance and the exact extent of the MBL phase in various quantum spin glass models \cite{PhysRevB.93.024202,PhysRevLett.113.200405,Mossi20160424,1742-5468-2017-12-123304} is of clear interest to quantum computing as it provides a theoretical tool for the analysis of how effectively quantum annealing (and other optimization methods that exploit the quantum dynamics of local Hamiltonians) explores the complicated energy landscapes of combinatorial problems.

MBL was proposed in \cite{Altshuler12446} as a physical mechanism behind the appearance of superexponentially closing gaps in the spectra of Hamiltonians in this later part of a quantum annealing run, with avoided crossing happening at a point that approaches the final time of the anneal, $~N^{-1/8}$. A more precise analysis \cite{RelevanceAvoidedCrossings}, refining the results of \cite{Altshuler12446}, and showed that these avoided crossings happen more realistically at a time that is $O(1)$ away from the end of the anneal. 

A further attempt to understand the dynamics of quantum spin glasses 
\cite{Smelyanskiy18_NonErgodic} considers a simplified ``impurity band'' model with a low-energy subspace composed of $M$ classical states whose energies are contained within a narrow interval, separated from the rest of the spectrum by an extensive energy gap (for simplicity it was assumed that the energies of states inside of the band are negative while all the others are zero). The dynamics of the system is defined by adding a transverse field term to the band Hamiltonian 
\begin{equation}
H = \sum_{i=1}^M E_i | i \rangle \langle i | -B_{\bot} \sum_{i=1}^N \sigma_i^x
\end{equation}
in order to induce transitions between different states. 
At $t=0$ the wavefunction of system is taken to be localized at one of the classical states $| i \rangle$ inside of the band. The dynamics generates multi-channel tunnelling from the initially populated site $i$ to other sites in the system. These tunnelling processes are mediated by the excited states that lie outside of the impurity band. This quantum process is expected to be hard to simulate classically; it is not captured by path-integral quantum Monte Carlo, which can be used to study tunnelling dynamics in potentials where tunnelling is dominated by a single path, such as the standard double-well potential \cite{Jiang17_Instanton,PhysRevLett.117.180402}.

After some time $t_{PT}$ the wavefunction will have spread to a superposition of some number $\Omega$ of states in the impurity band. This number of states turns out to be dependent on the fractal properties (controlled by the choice for $B_{\bot}$) of the eigenstates of the down-folded Hamiltonian that effectively describes the band. The best choice for $B_{\bot}$ is a compromise between two opposing considerations: ideally, one desires to tunnel to as many states inside of the impurity band as possible (and make $\Omega$ as large as possible), but one does not want to lose significant amplitude to excited states above the band. The optimal choice is achieved when the system is in a ``bad metal'' phase between the ``insulating'' (localized) phase and the ``conducting'' (ergodic, extendend) phase. In this phase, the eigenstates of the system are delocalized over a number of sites proportional to $N^{\alpha}$, for some value $0 < \alpha < 1$. Thus, the support of these states for all finite system sizes is not compact (that is, $O(1)$ in $N$, as in the case of localized states), but the fraction of sites in the support goes to zero in the thermodynamic limit (unlike ergodic extended states, where instead it approaches a finite value). States with these intermediate properties are called \emph{non-ergodic, extended} (NEE) states.

The typical time required to achieve this delocalization effect over a number $\Omega$ of states is given by the \emph{population transfer time} $t_{PT}$, which was found to be
\begin{equation}
t_{PT} \propto \Big(\frac{2^N}{\Omega \log \Omega} \Big)^{1/2} e^{n/(2B^2_{\bot})},
\end{equation}
which gives a scaling with $N$ and $\Omega$ that is comparable (up to a factor of $e^{n/(2B^2_{\bot})}$) to the scaling multiple-target Grover search, but is obtained through a different mechanism. The hope is that
investigating the dynamics of population transfer on more realistic examples of quantum spin glasses, and better understanding the role that non-ergodic, extended states play, will lead to a new class of quantum algorithms for optimization and sampling.

\section{Future Outlook}
\label{sec:future}


Since its birth as musings of Richard Feynman \cite{feynman1985quantum} 
and Yuri Manin \cite{manin2007mathematics}, 
quantum computing has blossomed from 
speculative idea into one with solid theoretical foundations, inspiring feats of engineering that are taking quantum computing into 
a new era. Many mysteries remain, from the fundamental to the practical.
Key questions include:
\begin{itemize}
    \item From which quantum effects does the power of quantum computing arise, and how best to harness these effects for computational purposes?
    \item How and when will quantum computers sufficiently large and robust to 
    support the solution of application-scale problems be built?
    \item How broad will the applications of quantum computing ultimately be?
 \end{itemize}
The emergence of quantum information processors large and coherent enough to be beyond the power of the world's largest supercomputers to simulate opens up new ways of working that will provide insight into the key questions in the field. How soon such processors arrive depend not only on hardware efforts, but also on the developing software ecosystem of algorithms and tools that make efficient use of quantum resources. Quantum computing work in the coming decades will require a different skill set and mindset that adds tool building and exploration of quantum algorithms through numerical experimentation to theoretical analysis that has been the predominant means of progress prior to the advent of quantum hardware of non-trivial size and coherence. 

With respect to the breadth of applications, it is instructive to look at the classical case. For the vast majority of computations run on a supercomputer, there is no mathematical proof that the algorithm being run is the best possible, or even a mathematical proof that it is better than last year's algorithm; the analysis of these algorithms on problem classes of interest is just too complex to carry out. Instead, algorithms are tested empirically, with practitioners comparing notes on their performance and formal competitions held, in SAT \cite{SAT_competition}, machine learning \cite{NIPS_competition}, and planning \cite{ICAPS_competition}, among others. Until recently this empirical approach was not possible for most quantum algorithms given the limited quantum hardware available and the exponential overhead in simulating the algorithms on supercomputers in the general case. The next decades, as the quantum hardware matures to support larger and more varied quantum algorithms, particularly quantum heuristic algorithms for which formal analysis is intractable on problem classes of interest, we strongly conjecture that the areas in which quantum computing is known to have an advantage will substantially broaden. After all, given the status in the classical case, what is the chance that most algorithms for which quantum computing provides a speed up are algorithms for which we can mathematically prove that they provide such a speed up? 
It seems  quantum computers only significantly outperform classical computers on problems for which the analysis is simple enough to provide a mathematical proof, given that this standard of proof is rare even for classical algorithms for most computationally challenging real-world  problems.

To understand the current status of the field, a historical perspective can be helpful. In 1972, under the guidance of Hans Mark, director of NASA Ames Research Center at the time, the  ``massively parallel'' Illiac IV \cite{barnes1968illiac,bouknight1972illiac} computer, was brought to Ames. It consisted of sixty-four 64-bit processors and, with a 50 MFLOP peak, was the fastest computer in the world at the time. Finding problems that demonstrated the potential of such parallel computers was challenging. It was not until 1975 that a usable service could be offered, but at a slower rate of operations \cite{hockney1988parallel}, and proponents had to continue to field questions as to whether such machines could ever compete with state-of-the-art approaches at the time, such as wind tunnels. The overwhelming progress of the succeeding decades makes it hard for us not to smile at these doubts in the early 70s. The NASA QuAIL team looks forward to working with other teams around the world to realize the potential of quantum computing building on our work to date from fundamental theory, through tools, to exploration of applications. If quantum computing has even a fraction of the success of massively parallel computers, its practical impact on the future will be great, with diverse applications including those beyond what we can imagine today.

\section*{Acknowledgements}
We are grateful to Bryan Biegel and Shon Grabbe for helpful comments on the manuscript, as well as to the anonymous reviewers. We are grateful for support from NASA Ames Research Center, NASA Advanced Exploration systems (AES) program, NASA Earth Science Technology Office (ESTO), and NASA Transformative Aeronautic Concepts Program (TACP). We also appreciate support from the AFRL Information Directorate under grant F4HBKC4162G001 and the Office of the Director of National Intelligence (ODNI) and the Intelligence Advanced Research Projects Activity (IARPA), via IAA 145483. D.V., Z.W., J.M. and S.H. are supported by NASA Academic Mission Services (NAMS), contract number NNA16BD14C. The views and conclusions contained herein are those of the authors and should not be interpreted as necessarily representing the official policies or endorsements, either expressed or implied, of ODNI, IARPA, AFRL, or the U.S. Government. The U.S. Government is authorized to reproduce and distribute reprints for Governmental purpose notwithstanding any copyright annotation thereon.

\section*{About the NASA QuAIL team}

The mandate of the NASA's Quantum Artificial Intelligence Laboratory (QuAIL) 
is to assess the potential impact of quantum computers on computational problems that will be faced by NASA missions in  decades to come. 
Successful and increasingly ambitious NASA missions require the solution of myriad challenging computational problems. QuAIL's home, the NASA Ames Research Center, has long had one of the most powerful supercomputers in the world.
This work requires highly interdisciplinary teams of theoretical and experimental physicists, computer scientists, and domain experts in the various application domains. The QuAIL team members, physicists, computer scientists, and mathematicians, come from a wide variety of backgrounds, with complementary expertise. The team has strong collaborations with application domain experts in and outside NASA, and with groups implementing quantum hardware, including 
Google and Rigetti with whom NASA has Space Act Agreements.
NASA's QuAIL team
has particular expertise in efficient
utilization of near-term quantum computing hardware to evaluate
the potential impact of quantum computing. 



\bibliographystyle{ieeetr} 

\bibliography{quail,extRefs}

\begin{thebibliography}{100}

\bibitem{RPbook}
E.~Rieffel and W.~Polak, {\em {Quantum Computing: A Gentle Introduction}}.
\newblock Cambridge, MA: MIT Press, 2011.

\bibitem{nielsen2010quantum}
M.~A. Nielsen and I.~L. Chuang, {\em Quantum Computation and Quantum
  Information}.
\newblock Cambridge University Press, 2010.

\bibitem{QAlgsZoo}
``{Quantum Algorithms Zoo }.'' http://quantumalgorithmzoo.org/.

\bibitem{shor1994algorithms}
P.~W. Shor, ``Algorithms for quantum computation: Discrete logarithms and
  factoring,'' in {\em Proceedings 35th annual {s}ymposium on {F}oundations of
  {C}omputer {s}cience}, pp.~124--134, Ieee, 1994.

\bibitem{preskill2018quantum}
J.~Preskill, ``Quantum computing in the {NISQ} era and beyond,'' {\em Quantum},
  vol.~2, p.~79, 2018.

\bibitem{boixo2018characterizing}
S.~Boixo, S.~V. Isakov, V.~N. Smelyanskiy, R.~Babbush, N.~Ding, Z.~Jiang, M.~J.
  Bremner, J.~M. Martinis, and H.~Neven, ``{Characterizing quantum supremacy in
  near-term devices},'' {\em Nature Physics}, vol.~14, pp.~1--6, jun 2018.

\bibitem{IBM}
IBM, ``{The IBM Quantum Experience}.'' http://www.research.ibm.com/quantum/,
  2016.

\bibitem{Sete16}
E.~A. Sete, W.~J. Zeng, and C.~T. Rigetti, ``{A functional architecture for
  scalable quantum computing},'' in {\em 2016 IEEE International Conference on
  Rebooting Computing, ICRC 2016 - Conference Proceedings}, 2016.

\bibitem{preskill2012quantum}
J.~Preskill, ``Quantum computing and the entanglement frontier,'' {\em
  arXiv:1203.5813}, 2012.

\bibitem{biswas2017nasa}
R.~Biswas, Z.~Jiang, K.~Kechezhi, S.~Knysh, S.~Mandra, B.~O’Gorman,
  A.~Perdomo-Ortiz, A.~Petukhov, J.~Realpe-G{\'o}mez, E.~Rieffel,
  D.~Venturelli, F.~Vasko, and Z.~Wang, ``A {NASA} perspective on quantum
  computing: Opportunities and challenges,'' {\em Parallel Computing}, vol.~64,
  pp.~81--98, 2017.

\bibitem{Ronnow2014defining}
{T. F. Ronnow}, {Z. Wang}, {J. Job}, {S. Boixo}, {S. V. Isakov}, {D. Wecker},
  {J. M. Martinis}, {D. A. Lidar}, and {M. Troyer}, ``{Defining and detecting
  quantum speedup},'' {\em Science}, vol.~345, no.~6195, pp.~420--424, 2014.

\bibitem{Boixo2014evidence}
S.~Boixo, T.~F. R{\o}nnow, S.~V. Isakov, Z.~Wang, D.~Wecker, D.~A. Lidar, J.~M.
  Martinis, and M.~Troyer, ``{Evidence for quantum annealing with more than one
  hundred qubits},'' {\em Nature Physics}, vol.~10, no.~3, pp.~218--224, 2014.

\bibitem{mandra2016strengths}
S.~Mandr{\`{a}}, Z.~Zhu, W.~Wang, A.~Perdomo-Ortiz, and H.~G. Katzgraber,
  ``{Strengths and weaknesses of weak-strong cluster problems: A detailed
  overview of state-of-the-art classical heuristics versus quantum
  approaches},'' {\em Physical Review A}, vol.~94, no.~2, p.~22337, 2016.

\bibitem{denchev2016computational}
V.~S. Denchev, S.~Boixo, S.~V. Isakov, N.~Ding, R.~Babbush, V.~Smelyanskiy,
  J.~Martinis, and H.~Neven, ``What is the computational value of finite-range
  tunneling?,'' {\em Physical Review X}, vol.~6, no.~3, p.~031015, 2016.

\bibitem{Farhi2014}
E.~Farhi, J.~Goldstone, and S.~Gutmann, ``A quantum approximate optimization
  algorithm,'' {\em arXiv:1411.4028}, 2014.

\bibitem{Wecker2016training}
D.~Wecker, M.~B. Hastings, and M.~Troyer, ``{Training a quantum optimizer},''
  {\em Physical Review A}, vol.~94, no.~2, p.~22309, 2016.

\bibitem{farhi2014quantum}
E.~Farhi, J.~Goldstone, and S.~Gutmann, ``{A Quantum Approximate Optimization
  Algorithm Applied to a Bounded Occurrence Constraint Problem},'' {\em
  arXiv:1412.6062}, 2014.

\bibitem{farhi2016quantum}
E.~Farhi and A.~W. Harrow, ``{Quantum Supremacy through the Quantum Approximate
  Optimization Algorithm},'' {\em arXiv:1602.07674}, 2016.

\bibitem{lin2016performance}
C.~Y.-Y. Lin and Y.~Zhu, ``{Performance of QAOA on Typical Instances of
  Constraint Satisfaction Problems with Bounded Degree},'' {\em
  arXiv:1601.01744}, 2016.

\bibitem{Jiang17_Grover}
Z.~Jiang, E.~G. Rieffel, and Z.~Wang, ``{Near-optimal quantum circuit for
  Grover's unstructured search using a transverse field},'' {\em Physical
  Review A}, vol.~95, no.~6, p.~62317, 2017.

\bibitem{guerreschi2017practical}
G.~G. Guerreschi and M.~Smelyanskiy, ``{Practical optimization for hybrid
  quantum-classical algorithms},'' {\em arXiv:1701.01450}, 2017.

\bibitem{Hadfield17_QAOAGen}
S.~Hadfield, Z.~Wang, B.~O'Gorman, E.~G. Rieffel, D.~Venturelli, and R.~Biswas,
  ``{From the quantum approximate optimization algorithm to a quantum
  alternating operator ansatz},'' {\em Algorithms}, vol.~12, no.~2, pp.~1--46,
  2019.

\bibitem{verdon2017quantum}
G.~Verdon, M.~Broughton, and J.~Biamonte, ``{A quantum algorithm to train
  neural networks using low-depth circuits},'' {\em arXiv:1712.05304}, 2017.

\bibitem{Wang18_QAOAFermionic}
Z.~Wang, S.~Hadfield, Z.~Jiang, and E.~G. Rieffel, ``{Quantum approximate
  optimization algorithm for MaxCut: A fermionic view},'' {\em Physical Review
  A}, vol.~97, p.~22304, feb 2018.

\bibitem{Hadfield17_QApprox}
S.~Hadfield, Z.~Wang, B.~O'Gorman, E.~Rieffel, D.~Venturelli, and R.~Biswas,
  ``{Quantum approximation with hard and soft constraints.},'' in {\em
  Proceedings of the Second International Workshop on Post Moore's Era
  Supercomputing}, pp.~15--21, 2017.

\bibitem{wang19XY}
Z.~Wang, N.~C. Rubin, J.~M. Dominy, and E.~G. Rieffel, ``${XY}$-mixers:
  analytical and numerical results for {QAOA},'' 2019.
\newblock arXiv:1904.09314.

\bibitem{grover1996fast}
L.~K. Grover, ``A fast quantum mechanical algorithm for database search,'' in
  {\em Proceedings 28th Annual {ACM} {S}ymposium on the {T}heory {O}f
  {C}omputing (STOC)}, pp.~212--219, ACM, 1996.

\bibitem{zalka_grovers_1999}
C.~Zalka, ``{Grover's quantum searching algorithm is optimal},'' {\em Physical
  Review A - Atomic, Molecular, and Optical Physics}, vol.~60, pp.~2746--2751,
  oct 1999.

\bibitem{Rabitz04}
H.~A. Rabitz, M.~M. Hsieh, and C.~M. Rosenthal, ``{Quantum Optimally Controlled
  Transition Landscapes},'' {\em Science}, vol.~303, no.~5666, pp.~1998--2001,
  2004.

\bibitem{Rabitz12}
R.~B. Wu, R.~Long, J.~Dominy, T.~S. Ho, and H.~Rabitz, ``{Singularities of
  quantum control landscapes},'' {\em Physical Review A - Atomic, Molecular,
  and Optical Physics}, vol.~86, p.~13405, jul 2012.

\bibitem{Rabitz16}
B.~Russell, H.~Rabitz, and R.~Wu, ``{Quantum Control Landscapes Are Almost
  Always Trap Free},'' {\em arXiv:1608.06198}, 2016.

\bibitem{Rabitz11}
R.~B. Wu, M.~A. Hsieh, and H.~Rabitz, ``{Role of controllability in optimizing
  quantum dynamics},'' {\em Physical Review A - Atomic, Molecular, and Optical
  Physics}, vol.~83, p.~62306, jun 2011.

\bibitem{mcclean2018barren}
J.~R. McClean, S.~Boixo, V.~N. Smelyanskiy, R.~Babbush, and H.~Neven, ``{Barren
  plateaus in quantum neural network training landscapes},'' {\em Nature
  communications}, vol.~9, no.~1, p.~4812, 2018.

\bibitem{Shabani16}
Z.~C. Yang, A.~Rahmani, A.~Shabani, H.~Neven, and C.~Chamon, ``{Optimizing
  variational quantum algorithms using pontryagin's minimum principle},'' {\em
  Physical Review X}, vol.~7, no.~2, 2017.

\bibitem{Nishimori98}
T.~Kadowaki and H.~Nishimori, ``{Quantum annealing in the transverse Ising
  model},'' {\em Physical Review E - Statistical Physics, Plasmas, Fluids, and
  Related Interdisciplinary Topics}, vol.~58, no.~5, pp.~5355--5363, 1998.

\bibitem{farhi2000quantum}
E.~Farhi, J.~Goldstone, S.~Gutmann, and M.~Sipser, ``{Quantum Computation by
  Adiabatic Evolution},'' {\em arXiv:quant-ph/0001106}, 2000.

\bibitem{Stollenwerk17}
T.~Stollenwerk, B.~O'Gorman, D.~Venturelli, S.~Mandr{\`a}, O.~Rodionova, H.~Ng,
  B.~Sridhar, E.~G. Rieffel, and R.~Biswas, ``Quantum annealing applied to
  de-conflicting optimal trajectories for air traffic management,'' {\em IEEE
  Transactions on Intelligent Transportation Systems}, 2019.

\bibitem{Tran16a}
T.~T. Tran, Z.~Wang, M.~Do, and E.~G. Rieffel, ``{Explorations of
  Quantum-Classical Approaches to Scheduling a Mars Lander Activity Problem},''
  in {\em Proceedings of the AAAI 2016 Workshop on Planning for Hybrid Systems
  (PlanHS-16)}, no.~1, pp.~641--649, 2016.

\bibitem{Tran16b}
T.~T. Tran, M.~Do, E.~G. Rieffel, J.~Frank, Z.~Wang, B.~O'Gorman,
  D.~Venturelli, and J.~C. Beck, ``{A Hybrid Quantum-Classical Approach to
  Solving Scheduling Problems},'' in {\em Ninth Annual Symposium on
  Combinatorial Search}, pp.~98--106, 2016.

\bibitem{Venturelli-JobShop-arXiv-2015}
D.~Venturelli, D.~J.~J. Marchand, and G.~Rojo, ``{Quantum Annealing
  Implementation of Job-Shop Scheduling},'' {\em arXiv:1506.08479}, 2015.

\bibitem{rieffel2015case}
V.~N. Smelyanskiy, M.~B. Do, D.~Venturelli, E.~G. Rieffel, E.~M. Prystay, and
  B.~O'Gorman, ``{A case study in programming a quantum annealer for hard
  operational planning problems},'' {\em Quantum Information Processing},
  vol.~14, no.~1, pp.~1--36, 2014.

\bibitem{OGorman2015}
B.~O. Gorman, E.~G. Rieffel, M.~Do, D.~Venturelli, and J.~Frank, ``{Compiling
  planning into quantum optimization problems : a comparative study},'' {\em
  Proceedings of the Workshop on Constraint Satisfaction Techniques for
  Planning and Scheduling Problems (COPLAS-15)}, pp.~11--20, 2014.

\bibitem{Perdomo17_Readiness}
A.~Perdomo-Ortiz, A.~Feldman, A.~Ozaeta, S.~V. Isakov, Z.~Zhu, B.~O'Gorman,
  H.~G. Katzgraber, A.~Diedrich, H.~Neven, J.~de~Kleer, B.~Lackey, and
  R.~Biswas, ``{On the readiness of quantum optimization machines for
  industrial applications},'' {\em arXiv:1708.09780}, aug 2017.

\bibitem{Benedetti17}
M.~Benedetti, J.~Realpe-G{\'{o}}mez, and A.~Perdomo-Ortiz, ``{Quantum-assisted
  Helmholtz machines: A quantum-classical deep learning framework for
  industrial datasets in near-term devices},'' {\em Quantum Science and
  Technology}, vol.~3, aug 2018.

\bibitem{Perdomo17_Opportunities}
A.~Perdomo-Ortiz, M.~Benedetti, J.~Realpe-G{\'o}mez, and R.~Biswas,
  ``Opportunities and challenges for quantum-assisted machine learning in
  near-term quantum computers,'' {\em Quantum Science and Technology}, vol.~3,
  no.~3, p.~030502, 2018.

\bibitem{Benedetti-2016}
M.~Benedetti, J.~Realpe-G{\'{o}}mez, R.~Biswas, and A.~Perdomo-Ortiz,
  ``{Estimation of effective temperatures in quantum annealers for sampling
  applications: A case study with possible applications in deep learning},''
  {\em Physical Review A}, vol.~94, no.~2, pp.~1--13, 2016.

\bibitem{benedetti2017quantum}
M.~Benedetti, J.~Realpe-G{\'{o}}mez, R.~Biswas, and A.~Perdomo-Ortiz,
  ``{Quantum-assisted learning of hardware-embedded probabilistic graphical
  models},'' {\em Physical Review X}, vol.~7, no.~4, p.~41052, 2017.

\bibitem{PerdomoOrtiz_EPJST2015}
A.~Perdomo-Ortiz, J.~Fluegemann, S.~Narasimhan, R.~Biswas, and V.~N.
  Smelyanskiy, ``{A quantum annealing approach for fault detection and
  diagnosis of graph-based systems},'' {\em European Physical Journal: Special
  Topics}, vol.~224, no.~1, pp.~131--148, 2015.

\bibitem{Jiang17_NonCommuting}
Z.~Jiang and E.~G. Rieffel, ``{Non-commuting two-local Hamiltonians for quantum
  error suppression},'' {\em Quantum Information Processing}, vol.~16, no.~4,
  p.~89, 2017.

\bibitem{marvian2017error}
M.~Marvian and D.~A. Lidar, ``{Error suppression for Hamiltonian quantum
  computing in Markovian environments},'' {\em Physical Review A}, vol.~95,
  no.~3, p.~32302, 2017.

\bibitem{wilson2018}
T.~H. Max~Wilson, Thomas~Vandal and E.~Rieffel, ``{Quantum-assisted associative
  adversarial network: Applying quantum annealing in deep learning},'' {\em
  arXiv:1904.10573}, 2019.

\bibitem{khoshaman2018quantum}
A.~Khoshaman, W.~Vinci, B.~Denis, E.~Andriyash, and M.~H. Amin, ``Quantum
  variational autoencoder,'' {\em Quantum Science and Technology}, vol.~4,
  no.~1, p.~014001, 2018.

\bibitem{benedetti2018quantum}
M.~Benedetti, J.~Realpe-G{\'o}mez, and A.~Perdomo-Ortiz, ``Quantum-assisted
  helmholtz machines: A quantum--classical deep learning framework for
  industrial datasets in near-term devices,'' {\em Quantum Science and
  Technology}, vol.~3, no.~3, p.~034007, 2018.

\bibitem{kingma2013auto}
D.~P. Kingma and M.~Welling, ``Auto-encoding variational bayes,'' {\em arXiv
  preprint arXiv:1312.6114}, 2013.

\bibitem{rolfe2016discrete}
J.~T. Rolfe, ``Discrete variational autoencoders,'' {\em arXiv preprint
  arXiv:1609.02200}, 2016.

\bibitem{khoshaman2018gumbolt}
A.~H. Khoshaman and M.~Amin, ``Gumbolt: Extending gumbel trick to boltzmann
  priors,'' in {\em Advances in Neural Information Processing Systems},
  pp.~4061--4070, 2018.

\bibitem{kingma2014semi}
D.~P. Kingma, S.~Mohamed, D.~J. Rezende, and M.~Welling, ``Semi-supervised
  learning with deep generative models,'' in {\em Advances in neural
  information processing systems}, pp.~3581--3589, 2014.

\bibitem{oord2016pixel}
A.~v.~d. Oord, N.~Kalchbrenner, and K.~Kavukcuoglu, ``Pixel recurrent neural
  networks,'' {\em arXiv preprint arXiv:1601.06759}, 2016.

\bibitem{salimans2017pixelcnn++}
T.~Salimans, A.~Karpathy, X.~Chen, and D.~P. Kingma, ``Pixelcnn++: Improving
  the pixelcnn with discretized logistic mixture likelihood and other
  modifications,'' {\em arXiv preprint arXiv:1701.05517}, 2017.

\bibitem{gulrajani2016pixelvae}
I.~Gulrajani, K.~Kumar, F.~Ahmed, A.~A. Taiga, F.~Visin, D.~Vazquez, and
  A.~Courville, ``Pixelvae: A latent variable model for natural images,'' {\em
  arXiv preprint arXiv:1611.05013}, 2016.

\bibitem{chen2016variational}
X.~Chen, D.~P. Kingma, T.~Salimans, Y.~Duan, P.~Dhariwal, J.~Schulman,
  I.~Sutskever, and P.~Abbeel, ``Variational lossy autoencoder,'' {\em arXiv
  preprint arXiv:1611.02731}, 2016.

\bibitem{sadeghi2018}
H.~Sadeghi, E.~Andriyash, W.~Vinci, L.~Buffoni, and M.~Amin, ``Pixelvae++:
  Improved pixelvae with discrete prior,'' in {\em Work in progress}.

\bibitem{vinci2019}
W.~Vinci, L.~Buffoni, H.~Sadeghi, E.~Andriyash, and M.~Amin, ``Training deep
  generative models with quantum annealers,'' in {\em Work in progress}.

\bibitem{feynman1982simulating}
P.~Binbaum and E.~Tromer, ``{Simulating physics with computers Presentation},''
  {\em Simulating physics with computers Presentation}, vol.~21, no.~6-7,
  pp.~467--488, 2005.

\bibitem{Abrams1997}
D.~S. Abrams and S.~Lloyd, ``{Simulation of many-body fermi systems on a
  universal quantum computer},'' {\em Physical Review Letters}, vol.~79,
  pp.~2586--2589, sep 1997.

\bibitem{Ortiz2001}
G.~Ortiz, J.~E. Gubernatis, E.~Knill, and R.~Laflamme, ``{Quantum algorithms
  for fermionic simulations},'' {\em Physical Review A - Atomic, Molecular, and
  Optical Physics}, vol.~64, no.~2, p.~14, 2001.

\bibitem{mcclean2016}
O.~Access, ``{The theory of variational hybrid quantum-classical algorithms
  Manuscript version : Accepted Manuscript},'' {\em New Journal of Physics},
  vol.~18, no.~2, p.~23023, 2016.

\bibitem{babbush2018low-depth}
R.~Babbush, N.~Wiebe, J.~McClean, J.~McClain, H.~Neven, and G.~K.~L. Chan,
  ``{Low-Depth Quantum Simulation of Materials},'' {\em Physical Review X},
  vol.~8, may 2018.

\bibitem{peruzzo2014}
A.~Peruzzo, J.~McClean, P.~Shadbolt, M.~H. Yung, X.~Q. Zhou, P.~J. Love,
  A.~Aspuru-Guzik, and J.~L. O'Brien, ``{A variational eigenvalue solver on a
  photonic quantum processor},'' {\em Nature Communications}, vol.~5, p.~4213,
  jul 2014.

\bibitem{kivlichan2018quantum}
I.~D. Kivlichan, J.~McClean, N.~Wiebe, C.~Gidney, A.~Aspuru-Guzik, G.~K.~L.
  Chan, and R.~Babbush, ``{Quantum Simulation of Electronic Structure with
  Linear Depth and Connectivity},'' {\em Physical Review Letters}, vol.~120,
  p.~110501, mar 2018.

\bibitem{guzik2005}
A.~Aspuru-Guzik, A.~D. Dutoi, P.~J. Love, and M.~Head-Gordon, ``{Chemistry:
  Simulated quantum computation of molecular energies},'' {\em Science},
  vol.~309, pp.~1704--1707, sep 2005.

\bibitem{Reiher2017}
M.~Reiher, N.~Wiebe, K.~M. Svore, D.~Wecker, and M.~Troyer, ``{Elucidating
  reaction mechanisms on quantum computers.},'' {\em Proceedings of the
  National Academy of Sciences of the United States of America}, vol.~114,
  no.~29, pp.~7555--7560, 2017.

\bibitem{drucker2009quantum}
A.~Drucker and R.~de~Wolf, ``Quantum proofs for classical theorems,'' {\em
  arXiv:0910.3376}, 2009.

\bibitem{Kuperberg99}
G.~Kuperberg, ``Random words, quantum statistics, central limits, random
  matrices,'' {\em Methods and Applications of Analysis}, vol.~9, no.~1,
  pp.~101--119, 2002.

\bibitem{barak2015beating}
B.~Barak, A.~Moitra, R.~O'Donnell, P.~Raghavendra, O.~Regev, D.~Steurer,
  L.~Trevisan, A.~Vijayaraghavan, D.~Witmer, and J.~Wright, ``Beating the
  random assignment on constraint satisfaction problems of bounded degree,''
  {\em arXiv:1505.03424}, 2015.

\bibitem{tang2018Aquantum}
E.~Tang, ``Quantum-inspired classical algorithms for principal component
  analysis and supervised clustering,'' {\em arXiv:1811.00414}, 2018.

\bibitem{tang2018Bquantum}
E.~Tang, ``A quantum-inspired classical algorithm for recommendation systems,''
  {\em arXiv:1807.04271}, 2018.

\bibitem{gilyen2018quantum}
A.~Gily{\'e}n, S.~Lloyd, and E.~Tang, ``Quantum-inspired low-rank stochastic
  regression with logarithmic dependence on the dimension,'' {\em arXiv
  preprint arXiv:1811.04909}, 2018.

\bibitem{vidal2004efficient}
G.~Vidal, ``Efficient simulation of one-dimensional quantum many-body
  systems,'' {\em Physical review letters}, vol.~93, no.~4, p.~040502, 2004.

\bibitem{tubman2011}
N.~M. Tubman, J.~L. Dubois, R.~Q. Hood, and B.~J. Alder, ``{Prospects for
  release-node quantum Monte Carlo},'' {\em Journal of Chemical Physics},
  vol.~135, no.~18, 2011.

\bibitem{tubman2012}
N.~M. Tubman and J.~McMinis, ``{Renyi Entanglement Entropy of Molecules:
  Interaction Effects and Signatures of Bonding},'' {\em arXiv:1204.4731},
  2012.

\bibitem{tubman2014}
N.~M. Tubman and D.~C. Yang, ``{Calculating the entanglement spectrum in
  quantum Monte Carlo with application to ab initio Hamiltonians},'' {\em
  Physical Review B - Condensed Matter and Materials Physics}, vol.~90,
  p.~81116, aug 2014.

\bibitem{tubman2016}
N.~M. Tubman, J.~Lee, T.~Y. Takeshita, M.~Head-Gordon, and K.~B. Whaley, ``{A
  deterministic alternative to the full configuration interaction quantum Monte
  Carlo method},'' {\em Journal of Chemical Physics}, vol.~145, no.~4, 2016.

\bibitem{qmcpack}
N.~Blunt, D.~C. Yang, M.~Casula, J.~Shea, T.~Beaudet, H.~Hao, J.~Vincent,
  C.~Bennett, W.~Parker, R.~Martin, A.~Tillack, P.~R.~C. Kent, B.~Rubenstein,
  E.~Neuscamman, b.~K. Clark, N.~Tubman, A.~Baczewski, I.~Kylanpaa,
  J.~Townsend, L.~Shulenburger, B.~van~der Goetz, A.~Mathuriya, S.~Zhang, M.~G.
  Lopez, D.~Ceperley, C.~Melton, Y.~Luo, K.~Esler, F.~Malone, M.~Dewing,
  J.~Kim, M.~Berrill, H.~Shin, Y.~Yang, Y.~W. Li, M.~A. Morales, O.~Heinonen,
  J.~McMinis, A.~Benali, S.~Chiesa, S.~Flores, E.~J.~L. Borda, K.~Delaney,
  J.~T. Krogel, N.~A. Romero, L.~Mitas, L.~Zhao, and R.~Clay, ``{QMCPACK : An
  open source ab initio Quantum Monte Carlo package for the electronic
  structure of atoms, molecules, and solids},'' {\em Journal of Physics:
  Condensed Matter}, vol.~30, no.~19, p.~195901, 2018.

\bibitem{tubman2015}
N.~M. Tubman, E.~Liberatore, C.~Pierleoni, M.~Holzmann, and D.~M. Ceperley,
  ``{Molecular-Atomic Transition along the Deuterium Hugoniot Curve with
  Coupled Electron-Ion Monte Carlo Simulations},'' {\em Physical Review
  Letters}, vol.~115, p.~45301, jul 2015.

\bibitem{Tubman2018}
N.~M. Tubman, C.~Mejuto-Zaera, J.~M. Epstein, D.~Hait, D.~S. Levine,
  W.~Huggins, Z.~Jiang, J.~R. McClean, R.~Babbush, M.~Head-Gordon, and K.~B.
  Whaley, ``{Postponing the orthogonality catastrophe: efficient state
  preparation for electronic structure simulations on quantum devices},'' {\em
  arXiv:1809.05523}, 2018.

\bibitem{hadfield2018divide}
S.~Hadfield and A.~Papageorgiou, ``{Divide and conquer approach to quantum
  Hamiltonian simulation},'' {\em New Journal of Physics}, vol.~20, no.~4,
  p.~43003, 2018.

\bibitem{yung2014transistor}
M.-H. Yung, J.~Casanova, A.~Mezzacapo, J.~McClean, L.~Lamata, A.~Aspuru-Guzik,
  and E.~Solano, ``{Trapped-ion computers for quantum chemistry.},'' {\em
  Scientific reports}, vol.~4, p.~3589, 2014.

\bibitem{OMalley2016}
P.~J. O'Malley, R.~Babbush, I.~D. Kivlichan, J.~Romero, J.~R. McClean,
  R.~Barends, J.~Kelly, P.~Roushan, A.~Tranter, N.~Ding, B.~Campbell, Y.~Chen,
  Z.~Chen, B.~Chiaro, A.~Dunsworth, A.~G. Fowler, E.~Jeffrey, E.~Lucero,
  A.~Megrant, J.~Y. Mutus, M.~Neeley, C.~Neill, C.~Quintana, D.~Sank,
  A.~Vainsencher, J.~Wenner, T.~C. White, P.~V. Coveney, P.~J. Love, H.~Neven,
  A.~Aspuru-Guzik, and J.~M. Martinis, ``{Scalable quantum simulation of
  molecular energies},'' {\em Physical Review X}, vol.~6, no.~3, p.~31007,
  2016.

\bibitem{Santagati2018}
R.~Santagati, J.~Wang, A.~A. Gentile, S.~Paesani, N.~Wiebe, J.~R. McClean,
  S.~R.~M. Short, P.~J. Shadbolt, D.~Bonneau, J.~W. Silverstone, D.~P. Tew,
  X.~Zhou, J.~L. OBrien, and M.~G. Thompson, ``{Witnessing eigenstates for
  quantum simulation of Hamiltonian spectra (Supplementary materials)},'' {\em
  Science Advances}, vol.~4, no.~1, p.~eaap9646, 2018.

\bibitem{Colless2018}
J.~I. Colless, V.~V. Ramasesh, D.~Dahlen, M.~S. Blok, J.~R. McClean, J.~Carter,
  W.~A. de~Jong, and I.~Siddiqi, ``{Robust determination of molecular spectra
  on a quantum processor},'' {\em Phys. Rev. X}, vol.~8, no.~1, p.~11021, 2017.

\bibitem{kandala2017}
J.~M. Gambetta, M.~Takita, K.~Temme, M.~Brink, A.~Kandala, A.~Mezzacapo, and
  J.~M. Chow, ``{Hardware-efficient variational quantum eigensolver for small
  molecules and quantum magnets},'' {\em Nature}, vol.~549, no.~7671,
  pp.~242--246, 2017.

\bibitem{wecker2015progress}
D.~Wecker, M.~B. Hastings, and M.~Troyer, ``{Towards Practical Quantum
  Variational Algorithms},'' {\em Phys. Rev. A}, vol.~92, no.~4, p.~042303,
  2015.

\bibitem{Rubin2018}
N.~C. Rubin, R.~Babbush, and J.~McClean, ``{Application of fermionic marginal
  constraints to hybrid quantum algorithms},'' {\em New Journal of Physics},
  vol.~20, no.~5, p.~53020, 2018.

\bibitem{mcclean2017hybrid}
J.~R. Mcclean, M.~E. Schwartz, J.~Carter, and W.~A.~D. Jong, ``{Hybrid
  Quantum-Classical Hierarchy for Mitigation of Decoherence and Determination
  of Excited States},'' {\em Phys. Rev. A}, vol.~95, no.~4, pp.~1--10, 2017.

\bibitem{Woitzik2018}
A.~Woitzik, ``Entanglement in quantum-classical variational algorithms,'' {\em
  Zulassungsarbeit, Albert-Ludwigs-Universit\"at Freiburg, Germany}, 2018.

\bibitem{Woitzik2019}
A.~Woitzik, P.~K. Barkoutsos, F.~Wudarski, C.~Fuchs, A.~Buchleitner, and
  I.~Tavernelli, ``Entanglement requirements for hybrid quantum-classical
  algorithms,'' {\em In preparation}, 2019.

\bibitem{huggins2018}
J.~Lee, W.~J. Huggins, M.~Head-gordon, and K.~Birgitta, ``Generalized unitary
  coupled cluster wavefunctions for quantum computation,'' {\em
  arXiv:1810.02327}, pp.~1--45, 2018.

\bibitem{Kitaev1995}
A.~Y. Kitaev, ``{Quantum measurements and the Abelian Stabilizer Problem},''
  {\em arXiv:quant-ph/9511026}, nov 1995.

\bibitem{Abrams1999}
D.~S. Abrams and S.~Lloyd, ``{Quantum algorithm providing exponential speed
  increase for finding eigenvalues and eigenvectors},'' {\em Physical Review
  Letters}, vol.~83, pp.~5162--5165, dec 1999.

\bibitem{Jones2012}
N.~C. Jones, J.~D. Whitfield, P.~L. McMahon, M.~H. Yung, R.~V. Meter,
  A.~Aspuru-Guzik, and Y.~Yamamoto, ``{Faster quantum chemistry simulation on
  fault-tolerant quantum computers},'' {\em New Journal of Physics}, vol.~14,
  no.~11, p.~115023, 2012.

\bibitem{wecker2014}
D.~Wecker, B.~Bauer, B.~K. Clark, M.~B. Hastings, and M.~Troyer, ``{Gate-count
  estimates for performing quantum chemistry on small quantum computers},''
  {\em Physical Review A - Atomic, Molecular, and Optical Physics}, vol.~90,
  p.~22305, aug 2014.

\bibitem{BabbushSpectra}
R.~Babbush, C.~Gidney, D.~W. Berry, N.~Wiebe, J.~McClean, A.~Paler, A.~Fowler,
  and H.~Neven, ``{Encoding Electronic Spectra in Quantum Circuits with Linear
  T Complexity},'' {\em Physical Review X}, vol.~8, p.~41015, oct 2018.

\bibitem{lanyon2010}
B.~P. Lanyon, J.~D. Whitfield, G.~G. Gillett, M.~E. Goggin, M.~P. Almeida,
  I.~Kassal, J.~D. Biamonte, M.~Mohseni, B.~J. Powell, M.~Barbieri,
  A.~Aspuru-Guzik, and A.~G. White, ``{Towards quantum chemistry on a quantum
  computer},'' {\em Nature Chemistry}, vol.~2, no.~2, pp.~106--111, 2010.

\bibitem{Wang2014}
Y.~Wang, F.~Dolde, J.~Biamonte, R.~Babbush, V.~Bergholm, S.~Yang, I.~Jakobi,
  P.~Neumann, A.~Aspuru-Guzik, J.~D. Whitfield, and J.~Wrachtrup, ``{Quantum
  Simulation of Helium Hydride Cation in a Solid-State Spin Register},'' {\em
  ACS Nano}, vol.~9, no.~8, pp.~7769--7774, 2015.

\bibitem{Neill17}
P.~Roushan, E.~Lucero, J.~M. Martinis, B.~Chiaro, A.~Megrant, K.~Kechedzhi,
  A.~Dunsworth, J.~Wenner, P.~Klimov, B.~Burkett, K.~Arya, A.~Vainsencher,
  J.~Mutus, H.~Neven, A.~Fowler, Z.~Chen, Y.~Chen, R.~Barends, S.~V. Isakov,
  M.~Giustina, T.~Huang, J.~Kelly, M.~Neeley, T.~C. White, S.~Boixo, D.~Sank,
  B.~Foxen, V.~Smelyanskiy, R.~Graff, E.~Jeffrey, C.~Quintana, and C.~Neill,
  ``{A blueprint for demonstrating quantum supremacy with superconducting
  qubits},'' {\em Science}, vol.~360, pp.~195--199, sep 2018.

\bibitem{harrow2017quantum}
A.~W. Harrow and A.~Montanaro, ``{Quantum computational supremacy},'' {\em
  Nature}, vol.~549, pp.~203--209, sep 2017.

\bibitem{bravyi2018quantum}
F.~L. Gall, ``{Average-Case Quantum Advantage with Shallow Circuits},'' {\em
  Science}, vol.~362, no.~6412, pp.~308--311, 2018.

\bibitem{markov2008simulating}
I.~L. Markov and Y.~Shi, ``{Simulating quantum computation by contracting
  tensor networks},'' {\em SIAM Journal on Computing}, vol.~38, no.~3,
  pp.~963--981, 2005.

\bibitem{bravyi2016improved}
S.~Bravyi and D.~Gosset, ``{Improved Classical Simulation of Quantum Circuits
  Dominated by Clifford Gates},'' {\em Physical Review Letters}, vol.~116,
  p.~250501, jun 2016.

\bibitem{villalonga2018}
B.~Villalonga, S.~Boixo, B.~Nelson, C.~Henze, E.~Rieffel, R.~Biswas, and
  S.~Mandr{\`{a}}, ``{A flexible high-performance simulator for the
  verification and benchmarking of quantum circuits implemented on real
  hardware},'' {\em arXiv:1811.09599}, 2018.

\bibitem{villalonga2019}
B.~Villalonga, D.~Lyakh, S.~Boixo, H.~Neven, T.~S. Humble, R.~Biswas, E.~G.
  Rieffel, A.~Ho, and S.~Mandr{\`a}, ``Establishing the quantum supremacy
  frontier with a 281 pflop/s simulation,'' {\em arXiv preprint
  arXiv:1905.00444}, 2019.

\bibitem{GoogleBristleconePreview}
{Julian K.}, ``{A Preview of Bristlecone, Google's New Quantum Processor}.''
  https://ai.googleblog.com/2018/03/a-preview-of-bristlecone-googles-new.html,
  2018.

\bibitem{note:db_release}
Numerical simulations available at
  https://data.nas.nasa.gov/quail/data.php?dir=/quaildata/quantum/qcSim.

\bibitem{ogorman2019parameterization}
B.~O'Gorman, ``Parameterization of tensor network contraction,'' in {\em 14th
  Conference on the Theory of Quantum Computation, Communication and
  Cryptography (TQC 2019)}, Leibniz International Proceedings in Informatics
  (LIPIcs), (Dagstuhl, Germany), Schloss Dagstuhl--Leibniz-Zentrum fuer
  Informatik, 2019.

\bibitem{Jiang17_Instanton}
Z.~Jiang, V.~N. Smelyanskiy, S.~V. Isakov, S.~Boixo, G.~Mazzola, M.~Troyer, and
  H.~Neven, ``{Scaling analysis and instantons for thermally assisted tunneling
  and quantum Monte Carlo simulations},'' {\em Physical Review A}, vol.~95,
  p.~12322, jan 2017.

\bibitem{Jiang17_PathIntegral}
Z.~Jiang, V.~N. Smelyanskiy, S.~Boixo, and H.~Neven, ``{Path-integral quantum
  Monte Carlo simulation with open-boundary conditions},'' {\em Physical Review
  A}, vol.~96, p.~42330, oct 2017.

\bibitem{kechedzhi2018efficient}
K.~Kechedzhi, V.~Smelyanskiy, J.~R. McClean, V.~S. Denchev, M.~Mohseni,
  S.~Isakov, S.~Boixo, B.~Altshuler, and H.~Neven, ``Efficient population
  transfer via non-ergodic extended states in quantum spin glass,'' in {\em
  13th Conference on the Theory of Quantum Computation, Communication and
  Cryptography}, 2018.

\bibitem{zhu2015efficient}
Z.~Zhu, A.~J. Ochoa, and H.~G. Katzgraber, ``Efficient cluster algorithm for
  spin glasses in any space dimension,'' {\em Physical review letters},
  vol.~115, no.~7, p.~077201, 2015.

\bibitem{Mandra17_Pitfalls}
S.~Mandr{\`{a}}, H.~G. Katzgraber, and C.~Thomas, ``{The pitfalls of planar
  spin-glass benchmarks: Raising the bar for quantum annealers (again)},'' {\em
  Quantum Science and Technology}, vol.~2, p.~38501, sep 2017.

\bibitem{mandra2018deceptive}
S.~Mandra and H.~G. Katzgraber, ``{A deceptive step towards quantum speedup
  detection},'' {\em Quantum Science and Technology}, vol.~3, no.~4, 2018.

\bibitem{Mandra17_Sampling}
S.~Mandr{\`{a}}, Z.~Zhu, and H.~G. Katzgraber, ``{Exponentially Biased
  Ground-State Sampling of Quantum Annealing Machines with Transverse-Field
  Driving Hamiltonians},'' {\em Physical Review Letters}, vol.~118, p.~70502,
  feb 2017.

\bibitem{ochoa2018feeding}
A.~J. Ochoa, D.~C. Jacob, S.~Mandr\`a, and H.~G. Katzgraber, ``Feeding the
  multitude: A polynomial-time algorithm to improve sampling,'' {\em Physical
  Review E}, vol.~99, p.~043306, apr 2019.

\bibitem{Rieffel14}
E.~G. Rieffel, D.~Venturelli, M.~Do, I.~Hen, and J.~Frank, ``{Parametrized
  Families of Hard Planning Problems from Phase Transitions},'' in {\em
  Proceedings of the Twenty-Eighth AAAI Conference on Artificial Intelligence},
  pp.~2337--2343, 2014.

\bibitem{Wang17_SMSPT}
Z.~Wang, B.~O. Gorman, T.~T. Tran, E.~G. Rieffel, J.~Frank, and M.~Do, ``{An
  Investigation of Phase Transitions in Single-Machine Scheduling Problems},''
  {\em Twenty-Seventh International Conference on Automated Planning and
  Scheduling}, no.~Icaps, pp.~325--329, 2017.

\bibitem{Wang17_Patch}
W.~Wang, S.~Mandr{\`{a}}, and H.~G. Katzgraber, ``{Patch-planting spin-glass
  solution for benchmarking},'' {\em Physical Review E}, vol.~96, p.~23312, aug
  2017.

\bibitem{marshall2018power}
J.~Marshall, D.~Venturelli, I.~Hen, and E.~G. Rieffel, ``Power of pausing:
  Advancing understanding of thermalization in experimental quantum
  annealers,'' {\em Phys. Rev. Applied}, vol.~11, p.~044083, Apr 2019.

\bibitem{Venturelli18_Compiling}
D.~Venturelli, M.~Do, E.~Rieffel, and J.~Frank, ``{Compiling quantum circuits
  to realistic hardware architectures using temporal planners},'' {\em Quantum
  Science and Technology}, vol.~3, p.~25004, apr 2018.

\bibitem{ghallab2004automated}
M.~Ghallab, D.~Nau, and P.~Traverso, {\em {Automated Planning: Theory and
  Practice}}.
\newblock Elsevier, 2004.

\bibitem{Booth18}
K.~E.~C. Booth, M.~Do, J.~C. Beck, E.~Rieffel, D.~Venturelli, and J.~Frank,
  ``{Comparing and Integrating Constraint Programming and Temporal Planning for
  Quantum Circuit Compilation},'' {\em arXiv:1803.06775}, mar 2018.

\bibitem{SPARK2019}
D.~Venturelli, M.~Do, B.~O’Gorman, J.~Frank, E.~Rieffel, K.~Booth, N.~Thanh,
  P.~Narayan, and S.~Nanda, ``Quantum circuit compilation: An emerging
  application for automated reasoning,'' {\em ICAPS SPARK Workshop (Open
  Review)}, 2019.

\bibitem{ogorman2019generalized}
B.~O'Gorman, W.~J. Huggins, E.~G. Rieffel, and K.~B. Whaley, ``Generalized swap
  networks for routing on nisq devices,'' 2019.

\bibitem{motta2018low-rank}
M.~Motta, E.~Ye, J.~R. McClean, Z.~Li, A.~J. Minnich, R.~Babbush, and G.~K.-L.
  Chan, ``{Low rank representations for quantum simulation of electronic
  structure},'' {\em arXiv:1808.02625}, 2018.

\bibitem{hadfield2016scientific}
M.~K. Bhaskar, S.~Hadfield, A.~Papageorgiou, and I.~Petras, ``{Quantum
  Algorithms and Circuits for Scientific Computing},'' {\em Quantum Information
  {\&} Computation}, vol.~16, no.~3-4, pp.~197--236, 2016.

\bibitem{hadfield2018boolean}
S.~Hadfield, ``{On the representation of Boolean and real functions as
  Hamiltonians for quantum computing},'' {\em arXiv:1804.09130}, 2018.

\bibitem{mandra2016faster}
S.~Mandr{\'{a}}, G.~G. Guerreschi, and A.~Aspuru-Guzik, ``{Faster than
  classical quantum algorithm for dense formulas of exact satisfiability and
  occupation problems},'' {\em New Journal of Physics}, vol.~18, no.~7,
  p.~73003, 2016.

\bibitem{Marshall17}
J.~Marshall, E.~G. Rieffel, and I.~Hen, ``{Thermalization, Freeze-out, and
  Noise: Deciphering Experimental Quantum Annealers},'' {\em Physical Review
  Applied}, vol.~8, p.~64025, dec 2017.

\bibitem{Smelyanskiy17_Environment}
V.~N. Smelyanskiy, D.~Venturelli, A.~Perdomo-Ortiz, S.~Knysh, and M.~I. Dykman,
  ``{Quantum Annealing via Environment-Mediated Quantum Diffusion},'' {\em
  Physical Review Letters}, vol.~118, p.~66802, feb 2017.

\bibitem{Kechedzhi16}
K.~Kechedzhi and V.~N. Smelyanskiy, ``{Open-system quantum annealing in
  mean-field models with exponential degeneracy},'' {\em Physical Review X},
  vol.~6, p.~21028, may 2016.

\bibitem{Venturelli15}
D.~Venturelli, S.~Mandr\`a, S.~Knysh, B.~O'Gorman, R.~Biswas, and
  V.~Smelyanskiy, ``Quantum optimization of fully connected spin glasses,''
  {\em Phys. Rev. X}, vol.~5, p.~031040, Sep 2015.

\bibitem{Smelyanskiy18_NonErgodic}
V.~N. Smelyanskiy, K.~Kechedzhi, S.~Boixo, S.~V. Isakov, H.~Neven, and
  B.~Altshuler, ``{Non-ergodic delocalized states for efficient population
  transfer within a narrow band of the energy landscape},'' {\em
  arXiv:1802.09542}, 2018.

\bibitem{d2016quantum}
L.~D'Alessio, Y.~Kafri, A.~Polkovnikov, and M.~Rigol, ``From quantum chaos and
  eigenstate thermalization to statistical mechanics and thermodynamics,'' {\em
  Advances in Physics}, vol.~65, no.~3, pp.~239--362, 2016.

\bibitem{huse2014phenomenology}
D.~A. Huse and V.~Oganesyan, ``{A phenomenology of certain many-body-localized
  systems},'' {\em Physical Review B - Condensed Matter and Materials Physics},
  vol.~90, p.~174202, nov 2013.

\bibitem{serbyn2013local}
M.~Serbyn, Z.~Papi{\'{c}}, and D.~A. Abanin, ``{Local conservation laws and the
  structure of the many-body localized states},'' {\em Physical Review
  Letters}, vol.~111, p.~127201, sep 2013.

\bibitem{ros2015integrals}
V.~Ros, M.~M{\"{u}}ller, and A.~Scardicchio, ``{Integrals of motion in the
  many-body localized phase},'' {\em Nuclear Physics B}, vol.~891,
  pp.~420--465, 2015.

\bibitem{basko2006metal}
D.~M. Basko, I.~L. Aleiner, and B.~L. Altshuler, ``{Metal-insulator transition
  in a weakly interacting many-electron system with localized single-particle
  states},'' {\em Annals of Physics}, vol.~321, pp.~1126--1205, may 2006.

\bibitem{imbrie2014many}
J.~Z. Imbrie, ``On many-body localization for quantum spin chains,'' {\em
  arXiv:1403.7837}, pp.~1--62, 2014.

\bibitem{imbrie_diagonalization_2016}
J.~Z. Imbrie, ``Diagonalization and many-body localization for a disordered
  quantum spin chain,'' {\em Physical Review Letters}, vol.~117, p.~27201, jul
  2016.

\bibitem{Altshuler12446}
B.~Altshuler, H.~Krovi, and J.~Roland, ``{Anderson localization makes adiabatic
  quantum optimization fail},'' {\em Proceedings of the National Academy of
  Sciences}, vol.~107, no.~28, pp.~12446--12450, 2010.

\bibitem{PhysRevB.93.024202}
C.~L. Baldwin, C.~R. Laumann, A.~Pal, and A.~Scardicchio, ``{The many-body
  localized phase of the quantum random energy model},'' {\em Physical Review
  B}, vol.~93, p.~24202, jan 2016.

\bibitem{PhysRevLett.113.200405}
C.~R. Laumann, A.~Pal, and A.~Scardicchio, ``{Many-body mobility edge in a
  mean-field quantum spin glass},'' {\em Physical Review Letters}, vol.~113,
  p.~200405, nov 2014.

\bibitem{Mossi20160424}
G.~Mossi and A.~Scardicchio, ``{Ergodic and localized regions in quantum spin
  glasses on the Bethe lattice},'' {\em Philosophical Transactions of the Royal
  Society A: Mathematical, Physical and Engineering Sciences}, vol.~375,
  no.~2108, 2017.

\bibitem{1742-5468-2017-12-123304}
C.~Monthus, ``{Random transverse field spin-glass model on the Cayley tree:
  Phase transition between the two many-body-localized phases},'' {\em Journal
  of Statistical Mechanics: Theory and Experiment}, vol.~2017, no.~12,
  p.~123304, 2017.

\bibitem{RelevanceAvoidedCrossings}
S.~Knysh and V.~Smelyanskiy, ``{On the relevance of avoided crossings away from
  quantum critical point to the complexity of quantum adiabatic algorithm},''
  2010.

\bibitem{PhysRevLett.117.180402}
S.~V. Isakov, G.~Mazzola, V.~N. Smelyanskiy, Z.~Jiang, S.~Boixo, H.~Neven, and
  M.~Troyer, ``{Understanding quantum tunneling through quantum Monte Carlo
  simulations},'' {\em Physical Review Letters}, vol.~117, p.~180402, oct 2016.

\bibitem{feynman1985quantum}
R.~P. Feynman, ``Quantum mechanical computers,'' {\em Optics news}, vol.~11,
  no.~2, pp.~11--20, 1985.

\bibitem{manin2007mathematics}
Y.~Manin, {\em {YURI I. MANIN. Mathematics as Metaphor: Selected Essays of Yuri
  I. Manin}}, vol.~17.
\newblock American Mathematical Soc., 2008.

\bibitem{SAT_competition}
``{SAT Competitions}.'' http://www.satcompetition.org/.

\bibitem{NIPS_competition}
``{NIPS Competition Track}.''
  http://https://nips.cc/Conferences/2018/CompetitionTrack/.

\bibitem{ICAPS_competition}
``{ICAPS Competitions}.''
  http://www.icaps-conference.org/index.php/Main/Competitions.

\bibitem{barnes1968illiac}
G.~Barnes, R.~Brown, M.~Kato, D.~Kuck, D.~Slotnick, and R.~Stokes, ``{The
  ILLIAC IV Computer},'' {\em IEEE Transactions on Computers}, vol.~C-17,
  no.~8, pp.~746--757, 2007.

\bibitem{bouknight1972illiac}
W.~Bouknight, S.~Denenberg, D.~McIntyre, J.~Randall, A.~Sameh, and D.~Slotnick,
  ``{The Illiac IV system},'' {\em Proceedings of the IEEE}, vol.~60, no.~4,
  pp.~369--388, 2008.

\bibitem{hockney1988parallel}
J.~Eastwood, {\em {Parallel computers: Architecture, programming and
  algorithms}}, vol.~27.
\newblock CRC Press, 2002.

\end{thebibliography}


\end{document}